\documentclass[journal, final, 10pt]{IEEEtran}
\usepackage[dvips]{graphics}
\usepackage[dvips]{changebar}
\usepackage{amsmath,bm}
\usepackage{amsfonts}
\usepackage{graphicx}
\usepackage{epsf, epsfig}
\usepackage{amssymb}

\ifCLASSINFOpdf
\else
\fi

\begin{document}
%
\title{\huge Non-Uniform Linear Antenna Array Design and Optimization for Millimeter Wave Communications}

%
%
%
\author{Peng~Wang,~\IEEEmembership{Member,~IEEE},
        Yonghui~Li,~\IEEEmembership{Senior~Member,~IEEE},
        Yuexing~Peng,~\IEEEmembership{Member,~IEEE},
        Soung Chang~Liew,~\IEEEmembership{Fellow,~IEEE},
        and~Branka~Vucetic,~\IEEEmembership{Fellow,~IEEE}
\thanks{This work was supported by the Australian Research Council (ARC) under Grants DP150104019, FT120100487, and by funding from the Faculty of Engineering \& Information Technologies, The University of Sydney, under the Faculty Research Cluster Program. The work of Y. Peng was supported by the National Natural Science Foundation of China under Grant 61171106, National Key Technology R\&D program of China with grant 2015ZX03002009-004, and the 863 Program with grant 2014AA01A705. The work of S. Liew was supported by the General Research Funds (Project No. 414713) established under the University Grant Committee of the Hong Kong Special Administrative Region, China.}
\thanks{Peng Wang is with Huawei Technologies, Sweden AB. His work in this paper was partially done when he was with the School of Electrical and Information Engineering, the University of Sydney, Sydney, Australia. E-mail: wp\_ady@hotmail.com.}
\thanks{Yonghui Li and Branka Vucetic are with the School of Electrical and Information Engineering, the University of Sydney, Sydney, Australia. E-mails: \{yonghui.li, branka.vucetic\}@sydney.edu.au.}
\thanks{Yuexing Peng is with the Key Lab of Universal Wireless Communication, Ministry of Education, Beijing University of Posts and Telecommunications, Beijing, 100876 China. E-mail: yxpeng@bupt.edu.cn.}
\thanks{Soung Chang~Liew is with the Department of Information Engineering, The Chinese University of Hong Kong, Hong Kong SAR, P. R. China. E-mail: soung@ie.cuhk.edu.hk.}
}

\maketitle

\begin{abstract}
    In this paper, we investigate the optimization of non-uniform linear antenna arrays (NULAs) for millimeter wave (mmWave) line-of-sight (LoS) multiple-input multiple-output (MIMO) channels. Our focus is on the maximization of the system \emph{effective multiplexing gain} (EMG), by optimizing the individual antenna positions in the transmit/receive NULAs. Here the EMG is defined as the number of signal streams that are practically supported by the channel at a finite SNR. We first derive analytical expressions for the asymptotic channel eigenvalues with arbitrarily deployed NULAs when, asymptotically, the end-to-end distance is sufficiently large compared to the aperture sizes of the transmit/receive NULAs. Based on the derived expressions, we prove that, the asymptotically optimal NULA deployment that maximizes the achievable EMG should follow the groupwise Fekete-point distribution. Specifically, the antennas should be physically grouped into $K$ separate uniform linear antenna arrays (ULAs) with the minimum feasible antenna spacing within each ULA, where $K$ is the target EMG to be achieved; in addition, the centers of these $K$ ULAs follow the Fekete-point distribution. We numerically verify the asymptotic optimality of such an NULA deployment and extend it to a \emph{groupwise projected arch type (PAT)} NULA deployment, which provides a more practical option for mmWave LoS MIMO systems with realistic non-asymptotic configurations. Numerical examples are provided to demonstrate a significant capacity gain of the optimized NULAs over traditional ULAs. 
\end{abstract}


%

\section{Introduction}
%
%
%
%

    \IEEEPARstart{W}{ith} the rapidly growing mobile services, there has been an ever increasing demand for very high wireless transmission data rates up to tens-of-Gigabits/second \cite{Cisco_14_19}\cite{Yoshi_TWC13}. The conventional microwave bands below 6 GHz have already been heavily utilized and cannot meet this demand \cite{Khan_Cam09}. Comparatively, the higher millimeter wave (mmWave) frequency band, ranging from 30GHz to 300GHz, offers large swathes of unlicensed spectrum and can potentially form the basis for the next revolution in wireless communications \cite{Peck_MMag09}-\cite{Peng_CMag15}.

    Although the mmWave band presents a very wide range of spectrum, it consists of many frequency segments with distinct channel characteristics and various service restrictions imposed by regulators in different countries \cite{Pi_CMag11}\cite{Yong_Eur07}. After excluding some sub-bands with severe atmospheric absorption that are unsuitable for outdoor wireless transmissions, the remaining segments are discretely distributed in the overall mmWave band. Aggregating these discrete bandwidth segments for mobile broadband communication remains a great challenge in the near future \cite{Peng_CMag15}. Currently, the widest commercially available single-channel mmWave bandwidth is 5 GHz, located at the E-band, ranging from 71-76 GHz and 81-86 GHz \cite{Peng_CMag15}\cite{FCC_05}\cite{Guo_EMC09}. Thus to support tens of gigabits/s transmission rates over a single mmWave channel with bandwidth no wider than 5 GHz, we must employ transmission schemes with very high spectral efficiencies (e.g., higher than 4 bits/s/Hz). However, due to very high operating frequencies, mmWave transceivers face new hardware design challenges such as increased phase noise, limited amplifier gain and the need for transmission line modeling of circuit components, which prevent the use of high order modulations in most mmWave transmission schemes \cite{Yong_Eur07}.

    Fortunately, thanks to the significantly reduced mmWave signal wavelengths, a large number of antennas can be packed into the transmitter/receiver with much smaller aperture sizes. This allows the use of multiple-input multiple-output (MIMO) techniques \cite{Telater_ETT99}\cite{Gold_JSAC03} to compensate for severe propagation losses of mmWave transmissions \cite{Yong_Eur07} and, at the same time, increase the system spectral efficiency by exploring the spatial domain. However,
    severe propagation loss also significantly reduces the richness of scattering in the mmWave propagation.
    When the transmitter and receiver are in view of each other without obstacles between them, the mmWave propagation is dominated by the line-of-sight (LoS) path. In this case, the channel is represented by a LoS MIMO model with highly correlated fading coefficients between different transmit-receive antenna pairs. Such a mmWave LoS MIMO channel matrix is typically rank deficient, which significantly degrades the achievable multiplexing gain of the channel \cite{Beach_URSI02}\cite{Araki_IEICE05}.

    There have been many research papers on 
    mmWave LoS MIMO systems. For example, for a point-to-point LoS MIMO channel with uniform linear antenna arrays (ULAs) at both link ends, \cite{Paulraj_TC02}-\cite{Nix_TVT07}
    showed that the channel vectors experienced by different transmit/receive antennas 
    can be mutually orthogonal if the antenna numbers and spacings of the transmit/receive ULAs as well as the communication distance between them satisfy the so-called \emph{Rayleigh distance criterion}, indicating that the maximum multiplexing gain is indeed achievable in pure LoS environments. In \cite{Peng_TWC14}, the authors considered a more practical scenario where the communication distance is larger than the Rayleigh distance. They showed that, in this case, the \emph{effective multiplexing gain} (EMG) of a ULA-based LoS MIMO channel that can be practically achieved is limited by the product of the aperture sizes of the transmit/receive ULAs. Here the EMG is defined as the number of spatially independent signal streams that can be practically supported by the channel at a finite signal-to-noise ratio (SNR). As a natural extension of ULAs, a nonuniform linear antenna array (NULA) allows its antenna elements to be non-uniformly distributed on a line segment, providing an additional dimension to optimize the EMG. However, the NULA deployment optimization problem is very complex analytically, and most existing papers on NULAs are based on simulations or brute force exhaustive search \cite{Mdhow_GlbC09}. How to systematically optimize the deployment of NULAs for maximizing the EMG is still an open problem.

    In this paper, we investigate the NULA deployment optimization for EMG maximization in mmWave LoS MIMO systems. Our contributions are as follows.
    \begin{itemize}
        \item[$\bullet$]    We derive analytical expressions for the eigenvalues of mmWave LoS channels with
        arbitrarily deployed transmit/receive NULAs when, asymptotically, the communication distance is sufficiently large compared to the aperture sizes of the transmit/receive NULAs;
        \item[$\bullet$]    Building on the asymptotic analysis, we analytically show that, the asymptotically optimal NULA deployment that maximizes the system EMG should be grouped into $K$ separate ULAs with the minimum feasible antenna spacing within each ULA, where $K$ is the target EMG to be achieved; in addition, the centers of these $K$ ULAs should follow the Fekete-point distribution \cite{Fekete_1923}. Such a deployment is referred to as the \emph{groupwise Fekete-point NULA deployment}. Its asymptotic optimality is verified via numerical examples;
        \item[$\bullet$]     We also investigate the NULA design in a non-asymptotic scenario through numerical optimizations. We show that the \emph{groupwise Projected arch type (PAT) NULA deployment}, which can be regarded as an extension of the groupwise Fekete-point one, is a suitable option for practical mmWave systems with non-asymptotic configurations. Several numerical examples are provided to demonstrate a performance gain obtained with optimized NULAs.
    \end{itemize}
    In summary, to the best of our knowledge, this is the first work that presents an analytical method to optimize the NULA deployment in mmWave LoS MIMO channels for EMG enhancement. The results provide useful insights into the design of tens-of-gigabits wireless communication systems over mmWave frequencies.

    \emph{Notations}: Boldface lower-case symbols represent vectors. Capital boldface characters denote matrices. The operators $(\cdot)^T$, $(\cdot)^H$ and $\|\cdot\|_2$ denote the transpose, conjugate-transpose and 2-norm of a matrix or vector, respectively. $\bm{I}_M$ represents an $M$-by-$M$ identity matrix. For a vector $\bm{a}$, $\text{diag}(\bm{a})$ is a diagonal matrix with $\bm{a}$ being the main diagonal.  For a square matrix $\bm{A}$, $\text{tr}(\bm{A})$ and $\text{det}(\bm{A})$ denote its trace and determinant, respectively. For an integer $N$, $\{1, 2, \cdots, N\}$ stands for the set consisting of $1, 2, \cdots, N$. For a set $\mathcal{S}$, $|\mathcal{S}|$ is the size of $\mathcal{S}$. 

\section{Preliminaries}
\subsection{System Model}
    Consider a fixed point-to-point mmWave LoS MIMO system with $N$ transmit and $M$ receive antennas. Assuming $M \leq N$, without loss of generality, and focusing on slowly varying frequency-flat fading channels, we model the transmission in the complex baseband as
    \begin{equation}
        \bm{r} = \bm{H}\bm{s} + \bm{n}
        \label{SystemModel}
    \end{equation}
    where $\bm{s} \in \mathbb{C}^{N \times 1}$ and $\bm{r} \in \mathbb{C}^{M \times 1}$ are, respectively, the transmitted and received signal vectors; $\bm{n} \in \mathbb{C}^{M \times 1}$ is a vector of independent and identically distributed (i.i.d.) complex additive white Gaussian noise (AWGN) samples with mean zero and variance $N_0$; and $\bm{H} = \{h_{m,n}\}\in \mathbb{C}^{M \times N}$ is the channel response matrix.  Since the pure LoS channel is considered here, the channel coefficient between each transmit-receive antenna pair is a deterministic function of the distance between them. Thus following the ray tracing principle, we model each entry of $\bm{H}$ as \cite{Eric_TWC11}
    \begin{eqnarray}
        h_{m,n} & = & \frac{\rho \lambda}{4 \pi d_{m,n}}e^{-j\frac{2\pi}{\lambda}d_{m,n}}, \ \ \forall m,n
        \label{ChannelModel1}
    \end{eqnarray}
    where $h_{m,n}$ is the channel coefficient from the $n$-th transmit antenna to the $m$-th receive antenna, $d_{m,n}$ is the distance between them, $\lambda$ is the signal wavelength, and $\rho$ contains all relevant constants such as attenuation and phase rotation caused by the antenna patterns at the transmitter/receiver.

    \begin{figure}[t]
        \centering
        \scalebox{0.6}{\includegraphics{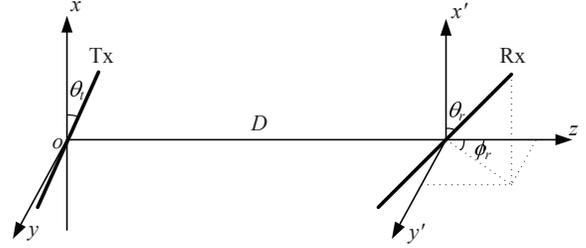}} \\
        \caption{The 3-D geometrical model for a mmWave LoS MIMO channel with arbitrarily deployed NULAs at both link ends.}
        \label{NULAmodel3D}
    \end{figure}

    Assume that two NULAs with aperture sizes $L_t$ and $L_r$ 
    are deployed at the transmitter and receiver, respectively, with arbitrary array orientations and antenna element distributions. We construct the following 3-D geometrical coordinate system to facilitate the calculation of $\{d_{m,n}\}$. As illustrated in Fig. \ref{NULAmodel3D}, the origin is located at the center of the transmit NULA. The $z$-axis is chosen as the line that connects the transmit NULA center and receive NULA center, pointing from the former to the latter. The $x$-axis is set such that the linear transmit NULA lies in the 2-D plane spanned by the $x$-axis and $z$-axis. Finally, the $y$-axis is determined by the right-hand rule based on the $x$-axis and $z$-axis. In this coordinate system, the receive NULA may have an arbitrary orientation. For specification, we use $\theta_t$ to represent the angle between the transmit NULA and the $x$-axis. In addition, we denote by $\theta_r$ the angle between the receive NULA and the $x$-axis, and by $\phi_r$ the angle between the projected vector of the receive NULA in the $y$-$z$ plane and the $z$-axis. Similar to those adopted in
    \cite{Oien_TWC07}-\cite{Peng_TWC14}\cite{Oien_TC09}-\cite{Oien_EJWCN},
    the 3-D geometrical model in Fig. \ref{NULAmodel3D} describes a communication system employing linear antenna arrays with arbitrary orientations.

    For the ease of describing the coordinates of the $n$-th transmit antenna, denoted by ($x_{t,n},y_{t,n},z_{t,n}$), we use $\alpha_{t,n} \in [-1, 1]$ to indicate its normalized position on the transmit NULA relative to the transmit NULA center. Then we have
    \begin{equation}
        \nonumber
        x_{t,n} = \frac{L_r\alpha_{t,n}\text{cos}\theta_t}{2}, \
        y_{t,n} = 0,   \ \text{and} \
        z_{t,n} = \frac{L_r\alpha_{t,n}\text{sin}\theta_t}{2}.
        \label{TxCoordinates}
    \end{equation}
    Similarly, let $\alpha_{r,m} \in [-1, 1]$ represent the normalized position of the $m$-th receive antenna on the receive NULA relative to its center. The coordinates of the $m$-th receive antenna relative to the center of the receive NULA, denoted by ($x_{r,m},y_{r,m},z_{r,m}$), are given by
    \begin{equation}
        \nonumber
        x_{r,m} = \frac{L_r\alpha_{r,m}\text{cos}\theta_r}{2}, \ y_{r,m} = \frac{L_r\alpha_{r,m}\text{sin}\theta_r\text{sin}\phi_r}{2}
    \end{equation}
    \begin{equation}
        \nonumber
        \mathrm{and} \ \  z_{r,m} = \frac{L_r\alpha_{r,m}\text{sin}\theta_r\text{cos}\phi_r}{2}.
    \end{equation}


    In addition, we assume a far-field communication distance throughout this paper \cite{Thiele_98}, i.e., the distance between the centers of the transmit and receive NULAs, denoted by $D$, is much larger than $L_t$ and $L_r$. Under this assumption, the path gains between all the transmit-receive antenna pairs are approximately the same and (\ref{ChannelModel1}) can be rewritten as
    \begin{equation}
        h_{m,n} \approx \frac{\rho \lambda}{4 \pi D}e^{-j\frac{2\pi}{\lambda}d_{m,n}}
        \label{ChannelModel1_Approx}
    \end{equation}
    with
    \begin{eqnarray}
        \nonumber
        d_{m,n} & \!\!\!\!=\!\!\!\!       & \sqrt{(x_{r,m} \!-\! x_{t,n})^2 \!+\! (y_{r,m} \!-\! y_{t,n})^2 \!+\! (D\!+\!z_{r,m} \!-\! z_{t,n})^2} \\
        \nonumber
        \ \     & \!\!\!\!\approx      \!\!\!\! & D + z_{r,m} + \frac{x_{r,m}^2 + y_{r,m}^2}{2D} - z_{t,n} + \frac{ x_{t,n}^2 + y_{t,n}^2 }{2D} \\
         & \!\!\!\!\      \!\!\!\! &- \frac{L_rL_t\text{cos}\theta_r\text{cos}\theta_t}{4D} \alpha_{r,m} \alpha_{t,n}.
        \label{DistanceApprox}
    \end{eqnarray}

    According to (\ref{ChannelModel1_Approx}) and (\ref{DistanceApprox}), we can decompose the channel matrix $\bm{H}$ as
    \begin{equation}
        \bm{H} = \frac{\rho \lambda}{4 \pi D} e^{-j\frac{2\pi D}{\lambda}}\bm{\Phi}_r \hat{\bm{H}} \bm{\Phi}_t
    \end{equation}
    where both $\bm{\Phi}_r \in  \mathbb{C}^{M \times M}$ and $\bm{\Phi}_t \in  \mathbb{C}^{N \times N}$ are diagonal matrices with their diagonal entries being $\{e^{-j\frac{2\pi}{\lambda}(z_{r,m} + (x_{r,m}^2 + y_{r,m}^2)/2D)} | m = 1, 2, \cdots, M\}$ and $\{e^{-j\frac{2\pi}{\lambda}(- z_{t,n} + (x_{t,n}^2 + y_{t,n}^2)/2D}) | n = 1, 2, \cdots, N\}$, respectively, and $\hat{\bm{H}} = \{\hat{h}_{m,n}\} \in  \mathbb{C}^{M \times N}$ is a full matrix with
    \begin{equation}
        \hat{h}_{m,n}\! =\! e^{j\frac{\pi L_rL_t\text{cos}\theta_r\text{cos}\theta_t}{2\lambda D} \alpha_{r,m} \alpha_{t,n}}  = e^{j\tau \alpha_{r,m} \alpha_{t,n}}
        \label{h_hat}
    \end{equation}
    and
    \begin{equation}
        \tau \doteq \frac{\pi L_rL_t\text{cos}\theta_r\text{cos}\theta_t}{2\lambda D}.
        \label{TauDefinition}
    \end{equation}
    Here, physically, $\tau$ represents the product of the effective transmit/receive NULA aperture sizes relative to the communication distance. Hence the value of $\tau$ reflects the range of the discrepancy of the channel coefficients of different transmit-receive antenna pairs. Since both $\bm{\Phi}_r$ and $\bm{\Phi}_t$ are unitary by definition, the singular values of $\bm{H}$ are identical to those of $\hat{\bm{H}}$ apart from a constant scaling factor $|\rho| \lambda/4 \pi D$.
    Define the channel gain matrix
    \begin{equation}
        \bm{G}_{M,N}(\tau) \doteq \hat{\bm{H}}\hat{\bm{H}}^H.
        \label{G_matrix}
    \end{equation}
    Denote by $\mu_{M, N}^{(m)}(\tau)$ the $m$-th largest eigenvalue of matrix $\bm{G}_{M,N}(\tau)$. In this paper, we will analyze the impact of antenna deployments, i.e., $\{\alpha_{r,m}\}$ and $\{\alpha_{t,n}\}$, on these eigenvalues $\{\mu_{M, N}^{(m)}(\tau) | m =1, 2, \cdots, M\}$ and optimize the individual positions of all transmit/receive antenna elements, i.e., $\{\alpha_{r,m}\}$ and $\{\alpha_{t,n}\}$, to improve the system performance.

\subsection{Uniform Linear Antenna Array and Rayleigh Distance}
    As a special case of NULA, a uniform linear antenna array (ULA) requires all the antenna elements to be equally spaced. In this case, we have
    \begin{equation}
        \alpha_{r,m} = \frac{2m-M-1}{M-1}, \forall m \ \text{and} \ \alpha_{t,n} = \frac{2n-N-1}{N-1}, \forall n.
    \end{equation}
    Consequently, matrix $\bm{G}_{M,N}(\tau)$ in (\ref{G_matrix}) can be further simplified, with its entries, denoted by $\{g_{m,n} |  m,n \in \{1, 2, \cdots, M\}\}$, expressed as \cite{Paulraj_TC02} $g_{m,n} = \text{sin}\frac{2\tau N (m-n)}{(M-1)(N-1)}/\text{sin}\frac{2\tau (m-n)}{(M-1)(N-1)}$.
    It has been shown in
    \cite{Paulraj_TC02}\cite{Oien_TWC07}\cite{Kruger_PIMRC03}
    that, at high SNRs, the maximum mutual information of the ULA-based mmWave LoS MIMO channel can be achieved when
    \begin{equation}
        D = D_{Ray}\text{cos}\theta_r\text{cos}\theta_t,
        \label{RayDistance}
    \end{equation}
    where $D_{Ray}  \doteq \frac{NL_rL_t}{\lambda(M-1)(N-1)}$ is called the Rayleigh distance\footnote{Here we have assumed $M \leq N$. Generally, the Rayleigh distance is defined as $D_{Ray} = \max\{M,N\}L_rL_t/\lambda(M-1)(N-1)$.}. Substituting (\ref{RayDistance}) into (\ref{TauDefinition}), we have $\tau = \frac{\pi(M-1)(N-1)}{2N}$ and $\bm{G}_{M,N}(\tau)$ reduces to a scaled identity matrix $ N \cdot \bm{I}_M$, indicating that the resultant channel 
    can support $M$ simultaneous spatial streams with equal channel quality \cite{Oien_TWC07}.

    It is also seen from (\ref{RayDistance}) that, when $D \leq D_{Ray}$, we can always find a proper ULA deployment for the angles $\theta_r$ and $\theta_t$ so that the Rayleigh distance criterion in (\ref{RayDistance}) is met. On the other hand, when $D > D_{Ray}$, it is not possible to satisfy (\ref{RayDistance}) unless we increase the aperture sizes of the transmit/receive ULAs  to make $D \leq D_{Ray}$. The resulting large aperture size of the transmitter/receiver is undesirable in practice. The scenario $D > D_{Ray}$ is most common in practical outdoor mmWave applications. For example, for a mmWave system with $\lambda = 0.004$m, $L_t = L_r = 0.6$m  and $M = N = 20$, from (\ref{RayDistance}) we have $D_{Ray} \approx 5$ m, which is far less than the expected outdoor mmWave communication distance (e.g., about 100 to 200 meters). This general scenario of $D > D_{Ray}$ was considered in \cite{Peng_TWC14}. It is shown in \cite{Peng_TWC14} that, though the system multiplexing gain (i.e., the rank of the channel matrix) may still remain unchanged, some channel eigenvalues vanish to zero quickly as $D$ increases and cannot be utilized for signal transmission when the system operating SNR is finite. Motivated by this, the authors in \cite{Peng_TWC14} introduced a more practical concept of \emph{effective multiplexing gain} (EMG) for the system, defined as
    \begin{equation}
        d_{M,N}(\tau) \doteq \sum\nolimits_{m=1}^M I\left(\mu_{M,N}^{(m)}(\tau)\big/\mu_{M,N}^{(1)}(\tau) \geq \Gamma \right).
        \label{EffectMG}
    \end{equation}
    In this definition, $I(\cdot)$ is an indicator function taking a value of 1 if its argument is true and 0 otherwise, and $\Gamma$ is a pre-determined threshold representing the minimum tolerable ratio between the qualities of those eigenmodes that are utilized for signal transmission. The higher the system operating SNR is, the less ratio between the qualities of the utilized eigenmodes can be tolerated, and in turn the less the value of the threshold $\Gamma$ can be set. When $\Gamma$ is properly chosen according to the system operating SNR, $d_{M,N}(\tau)$ corresponds to the number of independent spatial streams that can be practically supported by the channel. The main observation drawn in \cite{Peng_TWC14} is that, beyond the Rayleigh distance, the achievable EMG of the channel, $d_{M,N}(\tau)$, is limited by the product of the aperture sizes of the transmit and receive ULAs. Since the channel capacity increases linearly with the EMG, while logarithmically with the system SNR, i.e., $Cap \approx d_{M,N}(\tau)\mathrm{log}_2(1+\gamma)$ where $\gamma$ is the average SNR of each data stream, a higher EMG is preferable in practice. However, the only way to achieve a higher EMG in a ULA-based system is to increase the aperture sizes of the transmit and receive ULAs, $L_t$ and $L_r$, which is impractical in many systems. How to further maximize $d_{M,N}(\tau)$ in a more general NULA-based system with fixed $L_t$ and $L_r$ through the optimization of transmit/receive antenna element distributions is an open research problem, and it is the focus of this paper.

\section{Asymptotic Analysis}
    Before the NULA deployment optimization, we first consider the asymptotic channel characterization in the extreme of $\tau \to 0$, which will provide us with some insight of the channel behavior when $\tau$ is not zero but small, and facilitate our optimization problem formulation in the next section. Here $\tau \to 0$ corresponds to the case\footnote{Mathematically, $\tau \to 0$ also corresponds to the case of $\theta_t \to 0$ or $\theta_r \to 0$, where the system EMG can be simply improved through rotating the transmit or receive NULA. Hence this case is trivial and out of our interest in this paper.} in which the communication distance $D$ is sufficiently large compared to the product of the antenna aperture sizes of the transmit and receive NULAs, $L_t$ and $L_r$.

    For convenience, we define the following two Vandermonde matrices for a given integer $K$.
    \begin{equation}
        \bm{C}^{(r)}_{M \times K} = \left(
                    \begin{array}{ccccc}
                        1       & \alpha_{r,1}  & \alpha_{r,1}^2 & \cdots & \alpha_{r,1}^{K-1}  \\
                        1       & \alpha_{r,2}  & \alpha_{r,2}^2 & \cdots & \alpha_{r,2}^{K-1}  \\
                        \vdots  & \vdots        & \vdots         & \ddots & \vdots              \\
                        1       & \alpha_{r,M}  & \alpha_{r,M}^2 & \cdots & \alpha_{r,M}^{K-1}  \\
                    \end{array}
                \right)
        \label{Cr_matrix}
    \end{equation}
    and
    \begin{equation}
        \bm{C}^{(t)}_{N \times K} = \left(
                    \begin{array}{ccccc}
                        1       & \alpha_{t,1}  & \alpha_{t,1}^2 & \cdots & \alpha_{t,1}^{K-1}  \\
                        1       & \alpha_{t,2}  & \alpha_{t,2}^2 & \cdots & \alpha_{t,2}^{K-1}  \\
                        \vdots  & \vdots        & \vdots         & \ddots & \vdots              \\
                        1       & \alpha_{t,N}  & \alpha_{t,N}^2 & \cdots & \alpha_{t,N}^{K-1}  \\
                    \end{array}
                \right).
        \label{Ct_matrix}
    \end{equation}
    Let the QR decompositions of matrices $\bm{C}^{(r)}_{M \times M}$ and $\bm{C}^{(t)}_{N \times N}$ be, respectively,
    \begin{equation}
        \bm{C}^{(r)}_{M \times M} = \bm{Q}^{(r)}_M \bm{R}^{(r)}_{M \times M}
    \ \ \mathrm{and} \ \
        \bm{C}^{(t)}_{N \times N} = \bm{Q}^{(t)}_N \bm{R}^{(t)}_{N \times N},
        \label{QR_decomp_Tx}
    \end{equation}
    where $\bm{Q}^{(r)}_M$ ($\bm{Q}^{(t)}_N$) is an $M \times M$ ($N \times N$) unitary matrix, and $\bm{R}^{(r)}_{M \times M}$ ($\bm{R}^{(t)}_{N \times N}$) is an $M \times M$ ($N \times N$) upper triangular matrix. The following theorems show that the asymptotic behaviors of matrix $\bm{G}_{M,N}(\tau)$ are tractable.

    \emph{Theorem 1}: As $\tau \to 0$, the $m$-th largest eigenvalue of $\bm{G}_{M,N}(\tau)$ satisfies 
    \begin{eqnarray}
        \lim_{\tau \to 0} \frac{\text{ln}\mu_{M,N}^{(m)}(\tau)}{\text{ln}\tau} = 2(m-1), \forall m = 1, 2, \cdots, M.
        \label{Eigvalue_slope}
    \end{eqnarray}

    \emph{Theorem 2}: As $\tau \to 0$, the eigenvector of matrix $\bm{G}_{M,N}(\tau)$ corresponding to $\mu_{M,N}^{(m)}(\tau)$ converges to the $m$-th column of $\bm{Q}^{(r)}_M$ (see definition in 
    (\ref{QR_decomp_Tx})) for all $m = 1, 2,\cdots, M$.

    The proofs of Theorems 1 and 2 are given in Appendices A and B, respectively. From these two theorems, we can obtain the following corollary. Its proof is given in Appendix C.

    \emph{Corollary 1}: When $\tau$ is small, the $m$-th largest eigenvalue of matrix $\bm{G}_{M,N}(\tau)$ can be approximately represented as
    \begin{equation}
        \mu_{M,N}^{(m)}(\tau) \approx \left(\frac{r^{(r)}_m r^{(t)}_m}{(m-1)!}\right)^2 \tau^{2(m-1)} 
        \label{EigValueApprox}
    \end{equation}
    where $r^{(r)}_m$ ($r^{(t)}_m$) is the $m$-th diagonal entries of the matrices $\bm{R}_{M \times M}^{(r)}$ ($\bm{R}_{N \times N}^{(t)}$) defined in 
    (\ref{QR_decomp_Tx}).

\section{NULA Deployment Optimization}
    \subsection{Problem Formulation}
    Let us now return to the EMG maximization problem. 
    Denote by $\tau_{\text{min}}^{(K)}$ the critical value of $\tau$ when the step function $d_{M,N}(\tau)$ changes value from $K-1$ to $K$, i.e.,
    \begin{equation}
        \tau_{\text{min}}^{(K)} = \min \{\tau | d_{M,N}(\tau) = K \}.
        \label{Tau_min_K_definition}
    \end{equation}
    Physically, $\tau_{\text{min}}^{(K)}$ represents the minimum value of $\tau$ that can support an EMG of $K \ (K \leq M)$ at a practical SNR.
    Then from the property of non-decreasing step functions, we conclude that maximizing $d_{M,N}(\tau)$ for a given $\tau$ (beyond the Rayleigh distance) is equivalent to minimizing $\tau_{\text{min}}^{(K)}$ for a given $K$, and the EMG maximization problem can be initially formulated as
    \begin{equation}
        \textbf{P1:}   \ \ \ \ \ \ \ \ \ \ \ \ \ \ \ \ \ \min_{\{\alpha_{r,m}\}, \{\alpha_{t,n}\}} \tau_{\text{min}}^{(K)}.  \ \ \ \ \ \ \ \ \ \ \ \ \ \ \ \ \ \ \
    \end{equation}
    This is in line with the practical antenna design consideration, as a smaller $\tau_{\text{min}}^{(K)}$ corresponds to smaller transmit/receive aperture sizes $L_t, L_r$ and/or a longer communication distance $D$ that can achieve the same EMG of $K$, both of which are preferable for practical mmWave LoS MIMO systems. However, we are unable to express $\tau_{\text{min}}^{(K)}$ as an explicit function of $\{\alpha_{r,m}\}$ and $\{\alpha_{t,n}\}$ at present. Hence it is difficult to solve \textbf{P1} directly.

    In this paper, we consider the following alternative problem.
    \begin{equation}
        \textbf{P2:}   \ \ \ \ \ \ \ \ \ \ \ \ \ \max_{\{\alpha_{r,m}\}, \{\alpha_{t,n}\}} \prod\nolimits_{k=1}^{K}\mu_{M,N}^{(k)}(\tau),  \ \ \ \ \ \ \ \ \ \ \ \ \
    \end{equation}
   where $K$ is the target EMG\footnote{This target EMG should equal the maximum practically achievable EMG of the system. The latter can be easily determined via an exhaustive search. Hence we assume that the value of target EMG $K$ is known from now on.} to be achieved. Note that \textbf{P2} does not involve the threshold $\Gamma$ and so can be optimized independent of the value of $\Gamma$.

   We argue that \textbf{P2} is consistent with \textbf{P1} as follows. On one hand, recalling the definitions of $\tau_{\text{min}}^{(K)}$ and $d_{M,N}(\tau)$ in (\ref{Tau_min_K_definition}) and (\ref{EffectMG}), respectively, we can rewrite (\ref{Tau_min_K_definition}) as
    \begin{equation}
        \tau_{\text{min}}^{(K)} = \min \{\tau | \mu_{M,N}^{(K)}(\tau)/\mu_{M,N}^{(1)}(\tau) \geq \Gamma \}
    \end{equation}
    Hence the aim of \textbf{P1} is to minimize the value of $\tau$ under the constraint that the ratio between $\mu_{M,N}^{(K)}(\tau)$ and $\mu_{M,N}^{(1)}(\tau)$ is no less than a certain threshold $\Gamma$. This is equivalent to maximizing the ratio between $\mu_{M,N}^{(K)}(\tau)$ and $\mu_{M,N}^{(1)}(\tau)$ at a proper value of $\tau$.

    On the other hand, from (\ref{G_matrix}), we have the following natural constraint for \textbf{P2}.
   \begin{equation}
        \sum\nolimits_{k=1}^{K}\mu_{M,N}^{(k)}(\tau) \leq \sum\nolimits_{k=1}^{M}\mu_{M,N}^{(k)}(\tau) = \mathrm{tr}(\bm{G}_{M,N}(\tau))= MN.
        \label{Sum_mu1toK}
   \end{equation}
   According to the Cauchy-Schwarz inequality, for a set of non-negative variables $\{\mu_{M,N}^{(k)}(\tau)|k = 1, 2, \cdots, K\}$ with a finite upper bound $MN$ on their sum, their product is maximized when all these variables take the same value, and in turn the ratio between the minimum one (i.e, $\mu_{M,N}^{(K)}(\tau)$) and the maximum one (i.e., $\mu_{M,N}^{(1)}(\tau)$) is also maximized simultaneously. In other words, when the target function in \textbf{P2} is maximized, the ratio of $\mu_{M,N}^{(K)}(\tau)/\mu_{M,N}^{(1)}(\tau)$ will be maximized as well. Therefore, Problems \textbf{P1} and \textbf{P2} are consistent with each other in this sense.

%

    It is worth noting that \textbf{P2} is also consistent with a capacity maximization problem.
    Assume that we want to transmit $K \ (K \leq M)$ parallel data streams over the mmWave LoS MIMO channel in (\ref{SystemModel}). From the information theory for MIMO systems \cite{Tse_FWCbook}, the best way is to transmit them along the largest $K$ eigenmodes of the channel. Assuming a high transmit SNR $\gamma$ and equal power allocation among all the $K$ data streams, we write the corresponding channel capacity as
    \begin{eqnarray}
        \nonumber
        Cap & \!\!\!\!=\!\!\!\! & \sum_{k=1}^K \text{log}_2\left(1 \!\!+\!\! \frac{\gamma}{K}\mu_{M,N}^{(k)}(\tau)\right) \!\!\approx\!\! \sum_{k=1}^K \text{log}_2\left(\frac{\gamma}{K}\mu_{M,N}^{(k)}(\tau)\right)\\
        & \!\!\!\!=\!\!\!\! & K\text{log}_2(\gamma/K) + \text{log}_2\left(\prod_{k=1}^K \mu_{M,N}^{(k)}(\tau)\right).
        \label{ChCap_K}
    \end{eqnarray}
    Clearly, the first term in (\ref{ChCap_K}) is independent of the antenna deployment parameters $\{\alpha_{r,m}\}$ and $\{\alpha_{t,n}\}$, and the second term is consistent with \textbf{P2}, indicating that the solution to \textbf{P2} also leads to the maximization of the channel capacity in (\ref{ChCap_K}).

\subsection{Approximate NULA Deployment Optimization for Small $\tau$}
    Recall the approximate eigenvalue expression (\ref{EigValueApprox}) derived in Section III based on asymptotic analysis. When $\tau$ is small, we can substitute (16) into the target function of \textbf{P2} and rewrite it as
    \begin{eqnarray}
        \nonumber
        & \!\!\!\!\ \!\!\!\!& \prod_{k=1}^{K}\mu_{M,N}^{(k)}(\tau) \approx \prod_{k=1}^{K}\left(\frac{ r^{(r)}_k r^{(t)}_k}{(k-1)!}\right)^2 \tau^{2(k-1)} \\
        & \!\!\!\!=\!\!\!\! & \prod_{k=1}^{K}\left(\frac{\tau^{(k-1)}}{(k-1)!}\right)^2 \cdot \prod_{k=1}^{K}\left(r^{(r)}_k\right)^2 \cdot \prod_{k=1}^{K}\left(r^{(t)}_k\right)^2.
        \label{P2_Asy}
    \end{eqnarray}
    It is easy to see that the first multiplicative term in (\ref{P2_Asy}) is independent of the antenna deployment, while the second and third terms are determined by the transmit and receive antenna deployments $\{\alpha_{r,m}\}$ and $\{\alpha_{t,n}\}$, respectively. Hence to maximize (\ref{P2_Asy}) in \textbf{P2}, we only need to separately maximize the second and third terms in (\ref{P2_Asy}) via optimizing $\{\alpha_{r,m}\}$ and $\{\alpha_{t,n}\}$, respectively. Thus when $\tau$ is small, \textbf{P2} can be further decomposed into the following two problems.

    \textbf{Approximate NULA Deployment Optimization Criteria:}
    \begin{equation}
        \textbf{P3:}   \ \ \ \ \ \ \ \ \ \ \ \ \ \max_{-1 \leq \alpha_{r,1} \leq \alpha_{r,2} \leq \cdots \leq \alpha_{r,M} \leq 1} \prod_{k=1}^{K} \big(r^{(r)}_k\big)^2  \ \ \ \ \ \ \ \ \ \ \ \ \
        \label{RxOptProb}
    \end{equation}
    and
    \begin{equation}
        \textbf{P4:}   \ \ \ \ \ \ \ \ \ \ \ \ \ \max_{-1 \leq \alpha_{t,1} \leq \alpha_{t,2} \leq \cdots \leq \alpha_{t,N} \leq 1} \prod_{k=1}^{K} \big(r^{(t)}_k\big)^2. \ \ \ \ \ \ \ \ \ \ \ \
        \label{TxOptProb}
    \end{equation}
    By noting that \textbf{P3} and \textbf{P4} are very similar, our discussion below will be mainly focused on \textbf{P3}.

    To solve Problem \textbf{P3}, we need first to express each $r^{(r)}_k$ in (\ref{RxOptProb}) as an explicit function of $\{\alpha_{r,m}\}$. The following theorem provides a closed-form relationship between $r^{(r)}_k$ and $\{\alpha_{r,m}\}$.

    \emph{Theorem 3}: The diagonal entries of upper-triangular matrix $\bm{R}^{(r)}_{M \times M}$ in 
    (\ref{QR_decomp_Tx}) can be written as
    \begin{equation}
        r^{(r)}_1 = \sqrt{M}
        \label{R_diag_expression1}
    \end{equation}
    and\footnote{Note that when $k = 2$, each additive term in the denominator of (\ref{R_diag_expression2}) should be 1.}
    \begin{equation}
        r^{(r)}_k = \left(\frac{\sum\limits_{\overset{\mathcal{S} \subset \{1, 2, \cdots, M\},}{|\mathcal{S}| = k}}\prod\limits_{\overset{i<j,}{i,j\in \mathcal{S}}}\big(\alpha_{r,j} - \alpha_{r,i}\big)^2}{\sum\limits_{\overset{\mathcal{S} \subset \{1, 2, \cdots, M\},}{|\mathcal{S}| = k-1}}\prod\limits_{\overset{i<j,}{i,j\in \mathcal{S}}}\big(\alpha_{r,j} - \alpha_{r,i}\big)^2}\right)^{1/2}, \ \forall k>1.
        \label{R_diag_expression2}
    \end{equation}

    The proof of Theorem 3 can be found in Appendix D. According to Theorem 3, we have
    \begin{eqnarray}
        \nonumber
        \prod_{k=1}^K \big(r^{(r)}_k\big)^2
        & = & M\prod_{k=2}^K \frac{\sum\limits_{\overset{\mathcal{S} \subset \{1, 2, \cdots, M\},}{|\mathcal{S}| = k}}\prod\limits_{\overset{i<j,}{i,j\in \mathcal{S}}}\big(\alpha_{r,j} - \alpha_{r,i}\big)^2}{\sum\limits_{\overset{\mathcal{S} \subset \{1, 2, \cdots, M\},}{|\mathcal{S}| = k-1}}\prod\limits_{\overset{i<j,}{i,j\in \mathcal{S}}}\big(\alpha_{r,j} - \alpha_{r,i}\big)^2} \\
        & = & \sum\limits_{\overset{\mathcal{S} \subset \{1, 2, \cdots, M\},}{|\mathcal{S}| = K}}\prod\limits_{\overset{i<j,}{i,j\in \mathcal{S}}}\big(\alpha_{r,j} - \alpha_{r,i}\big)^2.
        \label{Prod_R_diag}
    \end{eqnarray}

    Substituting (\ref{Prod_R_diag}) into (\ref{RxOptProb}), we can reformulate \textbf{P3} as
    \begin{equation}
        \textbf{P5:}   \ \ \ \ \ \ \ \ \ \ \ \ \max_{-1 \leq \alpha_1 \leq \alpha_2 \leq \cdots \leq \alpha_M \leq 1} f_{M,K}(\bm{\alpha}) \ \ \ \ \ \ \ \ \ \
        \label{OptProb_new}
    \end{equation}
    where $\bm{\alpha} = (\alpha_1, \alpha_2, \cdots, \alpha_M)$,
    \begin{equation}
        f_{M,K}(\bm{\alpha}) \triangleq \sum\limits_{\overset{\mathcal{S} \subset \{1, 2, \cdots, M\},}{|\mathcal{S}| = K}}\prod\limits_{\overset{i<j,}{i,j\in \mathcal{S}}}\big(\alpha_j - \alpha_i\big)^2 \
        \label{f_MK}
    \end{equation}
    and the subscript $r$ has been omitted for brevity.

    \subsection{A Special Case: $K = M$}
    When $K = M$, the function $f_{M,K}(\bm{\alpha})$ reduces to
    \begin{equation}
         f_{K,K}(\bm{\alpha}) =  \prod\limits_{1 \leq i < j \leq K}\big(\alpha_j - \alpha_i\big)^2.
        \label{f_KK}
    \end{equation}
    It is easily seen that $f_{K,K}(\bm{\alpha})$ in (\ref{f_KK}) is just the squared determinant of the Vandermonde matrix constructed by $\{\alpha_1, \alpha_2, \cdots, \alpha_K\}$. Thus Problem \textbf{P5} (and in turn \textbf{P3}) reduces to the Vandermonde determinant maximization (VDM) problem \cite{VDM_JAT90} over the interval $[-1, +1]$. This kind of problems were first considered in \cite{Fekete_1923}\cite{Wely_1912} and the corresponding optimal values of $\{\alpha_k\}$, denoted by $\{\gamma_{K,k}|k = 1, 2, \cdots, K\}$, are referred to as Fekete points or Gauss-Lobatto points \cite{Bos_Math01}.

    \subsection{General Cases: $K \leq M$}
    The following theorem provides the optimal solution to Problem \textbf{P5} in the general case\footnote{    We numerically find that, even if $K$ does not divide $M$, the NULA deployment in (\ref{Opt_alpha}) is still optimal. However, we are unable to prove it at present.} of $K \leq M$ when $K$ divides $M$. Its proof can be found in Appendix E

    \emph{Theorem 4}: When $K$ divides $M$, the optimal solution to \textbf{P5} is to divide $\{\alpha_m |m = 1, \cdots, M\}$ into $K$ equal-size groups, and let all $\{\alpha_m\}$ in the $k$-th group take the same value 
    of $\gamma_{K,k}$, i.e.,
    \begin{equation}
        \alpha_m = \gamma_{K,k},\ \text{if} \  k-1 < mK/M \leq k.
        \label{Opt_alpha}
    \end{equation}


    In summary, we should divide all the $M$ antenna elements into $K$ groups with approximately the same sizes. The antennas in the same group should be compactly co-located, e.g., forming a ULA with the minimum spacing of $\lambda/2$, and the centers of these $K$ groups should follow the above-mentioned Fekete-point distribution. This groupwise deployment can be intuitively understood as follows. Since we aim to achieve an EMG of $K$, only $K$ distinct eigenmodes are required to support $K$ spatially independent signal streams, and the rest eigenmodes are unnecessary. Thus by dividing all the antennas into $K$ compact groups, we can already guarantee $K$ distinct eigenmodes. The antennas in the same group can be completely utilized to provide power gain for a capacity enhancement. Note that the conclusion of letting all group centers to follow the Fekete-point distribution is drawn in the extreme case of $\tau \to 0$. Therefore, we can only guarantee its optimality in this asymptotic case. It may not be optimal in the non-asymptotic case when $\tau$ takes finite values. We will discuss this practical scenario in Section V.

\subsection{Fekete-Point Distribution}
    \begin{figure}[t]
        \centering
        \scalebox{0.43}{\includegraphics{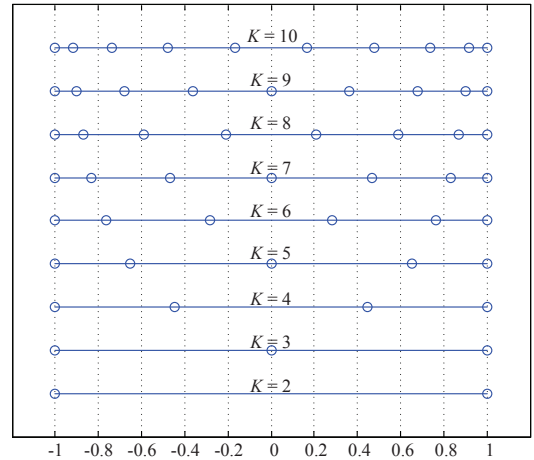}} \\
        \caption{Illustration of Fekete points for $K = 2, 3, \cdots, 10$.}
        \label{FeketePoint2to10_Fig}
    \end{figure}

    Till now, we have analytically shown that the asymptotically optimal NULA deployment is closely related to the Fekete-point distribution. It is well known that finding the exact values of all Fekete points within a general compact set\footnote{A set $\mathcal{S}$ is compact if for every open cover of $\mathcal{S}$ there exists a finite subcover of $\mathcal{S}$.} is a difficult and open problem (Problem 7 of \cite{Smale_Math98}). However, when the compact set reduces to the one-dimensional interval $[-1, 1]$, we can readily show the following property for the function (\ref{f_KK}).

    \emph{Property 1:} The function $f_{K,K}(\bm{\alpha})$ in (\ref{f_KK}) is strictly quasi-convex over the set of $\mathcal{S}_\alpha \triangleq \{(\alpha_1, \alpha_2, \cdots, \alpha_K)|-1 = \alpha_1 \leq \alpha_2 \leq \cdots \leq \alpha_K = 1\}$.

%

    Property 1 can be proved by directly checking if $f_{K,K}(\bm{\alpha})$ satisfies the definition of the strictly quasi-convex function \cite{ConOpt_Boyd04}.
    Due to space limitation, we skip the proof here. Based on Property 1, we can adopt the standard steepest descend method with adaptive step length \cite{ConOpt_Boyd04} to find the corresponding Fekete points that maximize $f_{K,K}(\bm{\alpha})$. The details are omitted here for brevity.

    Fig. \ref{FeketePoint2to10_Fig} illustrates some Fekete-point distributions for $K$ being up to 10. Their specific values are listed in Table \ref{FeketePoints2to10_Tab}. We can see that when $K = 2$ and 3, the Fekete-point distribution reduces to the conventional uniform distribution. While when $K \geq 4$, the Fekete-point distribution distinguishes itself from the uniform one by ``pushing'' the points towards the two ends of the interval and exhibits a symmetric and centrifugal distribution.

\subsection{Projected Arch Type (PAT) Distribution}
    According to the symmetric and centrifugal property of the Fekete-point distribution shown above, we develop the following projected arch type (PAT) distribution to approximate the Fekete-point distribution, which leads to an extension of the groupwise Fekete-point antenna deployment and will facilitate a practical implementation of the latter.

    \begin{table}
        \centering
        \caption{Detailed Values of Fekete Points with $K = 2, 3, \cdots, 10$}
    \begin{tabular}{|l|l|}
        \hline
        $K$     &    $\ \ \ \ \ \ \ \ \ \ \ \ \ \ \ \ \ \ \ \ \ \ \ \ \ \ \ \ \  $ Fekete points  \\
        \hline
        \hline
        2       & $-1, 1$  \\
        \hline
        3       & $-1, 0, 1$   \\
        \hline
        4       & $-1, -0.4472, 0.4472, 1$   \\
        \hline
        5       & $-1, -0.6547, 0, 0.6547, 1$   \\
        \hline
        6       & $-1, -0.7651, -0.2852, 0.2852, 0.7651, 1$   \\
        \hline
        7       & $-1, -0.8302, -0.4688, 0, 0.4688, 0.8302, 1$   \\
        \hline
        8       & $-1, -0.8717, -0.5917, -0.2093, 0.2093, 0.5917, 0.8717, 1$   \\
        \hline
        9       & $-1, -0.8998, -0.6772, -0.3631, 0, 0.3631, 0.6772, 0.8998, 1$   \\
        \hline
        10      & $-1, -0.9195, -0.7388, -0.4779, -0.1653, 0.1653, 0.4779, $   \\
        \       & $0.7388, 0.9195, 1$ \\
        \hline
    \end{tabular}
        \label{FeketePoints2to10_Tab}
    \end{table}

    \begin{figure}[t]
        \centering
        \scalebox{0.43}{\includegraphics{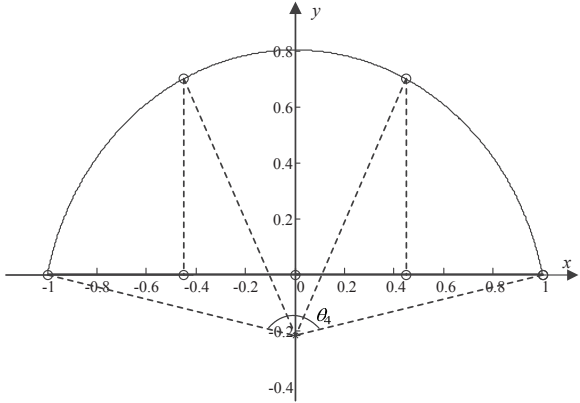}} \\
        \caption{Illustration of the approximation of the Fekete points using the projected arch type distribution with $K = 4$.}
        \label{ProjectArchType_Approx4_Fig}
    \end{figure}

    Let us take the case of $K = 4$ as an example. As shown in Fig. \ref{ProjectArchType_Approx4_Fig}, we consider an arch with a certain angle $\theta$. The 2-D coordinate system is constructed by letting the chord corresponding to this arch be on the $x$-axis with its center point located at the origin. For convenience, we further normalize the length of the chord to be 2. Then we uniformly distribute $K = 4$ points on the arch. By projecting these four points onto the $x$-axis, we can obtain a symmetric and centrifugal 4-point distribution, which we refer to as the projected arch type (PAT) distribution. It can be expected that, when the value of $\theta$ is properly chosen, we can generate a good approximation for the Fekete-point distribution with $K = 4$. In addition, such an approximated distribution can be characterized by a single parameter $\theta$. Mathematically, we can approximate all the Fekete points $\{\gamma_{K,k}|k = 1, 2, \cdots, K\}$ as
    \begin{equation}
        \gamma_{K,k} \approx \widetilde{\gamma}_{K,k} \triangleq \frac{\sin \frac{(2k-1-K)\theta_K}{2(K-1)}}{\sin \frac{\theta_K}{2}}
        \label{FeketePoint_Approx}
    \end{equation}
    where $\theta_K$ is the optimized value of $\theta$ that minimizes the approximation error, i.e.,
        $\theta_K = \arg\min_{\theta} \|\bm{\gamma}_K - \widetilde{\bm{\gamma}}_K\|_2$
    and $\widetilde{\bm{\gamma}}_K = (\widetilde{\gamma}_{K,1}\ \widetilde{\gamma}_{K,2}\ \cdots \ \widetilde{\gamma}_{K,K})$.

    \begin{table}
        \centering
        \caption{Values of $\theta_K$ for PAT Approximation and Their Approximation Error}
    \begin{tabular}{|c|c|c|}
        \hline
        $K$     &   $\theta_K$  & $\|\bm{\gamma}_K - \widetilde{\bm{\gamma}}_K\|$ \\
        \hline
        \hline
         4      &   2.7136      & 0 \\
        \hline
         5      &   2.8066      & 0 \\
         \hline
         6      &   2.8660      & $2.689\times 10^{-4}$ \\
         \hline
         7      &   2.9074      & $3.3458\times 10^{-4}$ \\
         \hline
         8      &   2.9378      & $3.5097\times 10^{-4}$ \\
         \hline
         9      &   2.9612      & $3.4593\times 10^{-4}$ \\
         \hline
         10     &   2.9798      & $3.3158\times 10^{-4}$ \\
        \hline
    \end{tabular}
        \label{PATApprox4to10_Tab}
    \end{table}

    Table \ref{PATApprox4to10_Tab} lists the values of $\{\theta_K\}$ and their corresponding approximation errors. We can see that when $K = 4 \ \text{and} \  5$, the Fekete points can be exactly described using the PAT distribution, and when $K > 5$, the PAT approximation errors are only of the order $10^{-4}$. This indicates that our PAT approximation is very accurate. Note that we did not include the cases of $K = 2$ and 3 as their corresponding PAT approximation is always exactly accurate for any $\theta$.

%

\section{Numerical Examples and Discussions}
    Till now, we have analytically optimized the NULA deployment in the extreme case of $\tau \to 0$. In this section, We numerically validate the asymptotic optimality of the proposed NULA deployment. We also verify if such an NULA deployment can improve the capacity over conventional ULAs in practical mmWave environments with non-vanishing $\tau$ (i.e., with a finite communication distance and non-vanishing transmit/receive NULA aperture sizes). In addition, we propose a groupwise PAT NULA deployment as an extension of the asymptotically optimal groupwise Fekete-point NULA deployment for practical mmWave LoS MIMO systems.

\subsection{Some Numerical Examples: A First Glance}

    \begin{figure}[t]
        \centering
        \scalebox{0.43}{\includegraphics{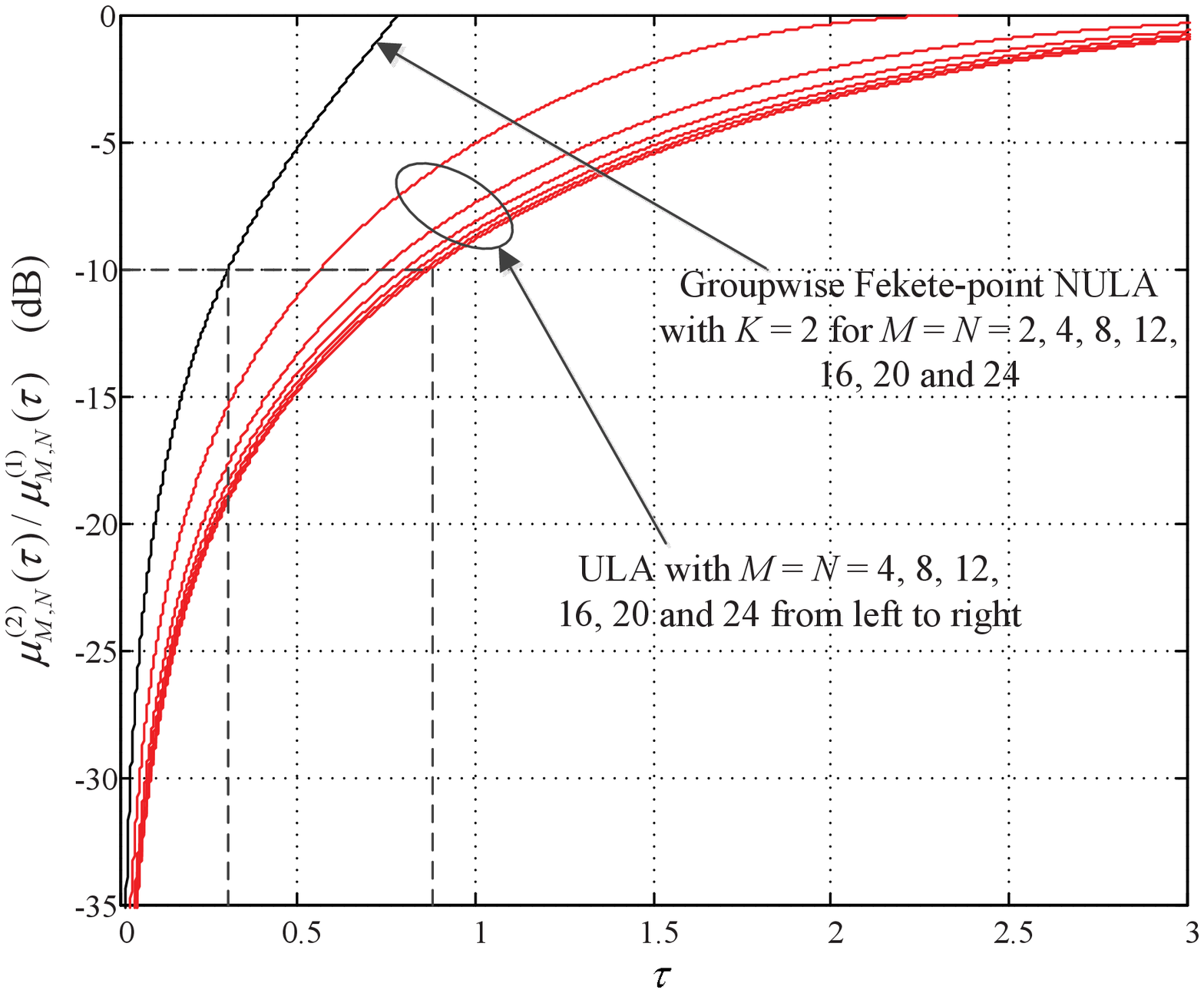}} \\
        \scalebox{0.75}{\includegraphics{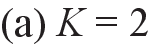}} \\
        \scalebox{0.43}{\includegraphics{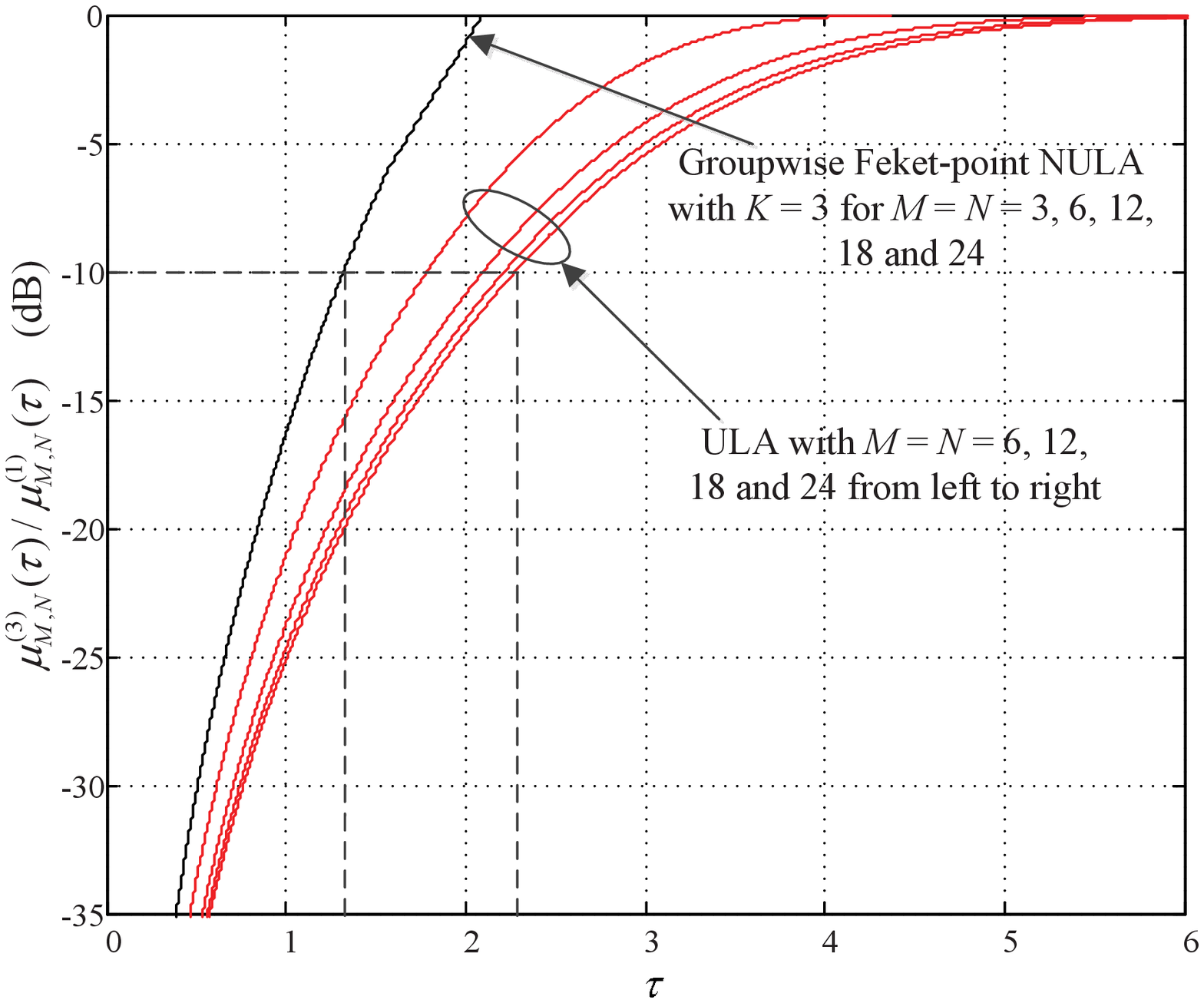}} \\
        \scalebox{0.75}{\includegraphics{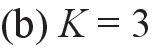}} \\
        \caption{The values of $\mu_{M,N}^{(K)}(\tau)/\mu_{M,N}^{(1)}(\tau)$ versus $\tau$ achieved by ULAs and optimized NULAs in mmWave LoS MIMO channels with variable numbers of transmit and receive antennas and (a) $K = 2$ and (b) $K = 3$, respectively.}
        \label{Tau_vs_mu2and3overmu1_Fig}
    \end{figure}

    We first consider the system designs that achieve the EMG up to 3. Fig. \ref{Tau_vs_mu2and3overmu1_Fig} plots the curves of $\mu_{M,N}^{(K)}(\tau)/\mu_{M,N}^{(1)}(\tau)$ ($K=2$ in (a) and $K = 3$ in (b)) versus $\tau$ achieved by ULAs and optimized NULAs in mmWave LoS MIMO channels with variable numbers of transmit/receive antennas up to 24. From Fig. \ref{Tau_vs_mu2and3overmu1_Fig}, we can see that when $\Gamma = -10$dB, the value of $\tau_{\min}^{(2)}$ achieved in the ULA-based system with $M = N = 24$ is 0.8776. For comparison, the corresponding optimized NULA-based system has the $\tau_{\min}^{(2)}$ value of only 0.3063. Similarly, from Fig. \ref{Tau_vs_mu2and3overmu1_Fig}(b) we can obtain the value of $\tau_{\min}^{(3)}$ to be 2.2821 and 1.3218, respectively, for the ULA-based system with $M = N = 24$ and the corresponding optimized NULA-based system. This indicates that to maintain the same EMG of $K = 2$ or 3, the latter system can communicate over a longer distance, or requires less transmitter/receiver aperture sizes, than the former system. For example, if such a system lies indoor operating at 60 GHz with transmit/receive NULA aperture sizes $L_t = L_r = 0.1$ meter, the conventional ULA-based antenna deployment can only maintain an EMG of 2 and 3 up to a communication distance of
    \begin{equation}
        D = \frac{\pi L_t L_r}{2\lambda \tau_{\min}^{(2)}} \approx 3.58 \ \text{meters} \ \ \mathrm{and} \ \ D = \frac{\pi L_t L_r}{2\lambda \tau_{\min}^{(3)}} \approx 1.38 \ \text{meters},
    \end{equation}
    respectively, while with our optimized NULA deployment, these distances can be increased to 10.26 meters and 2.38 meters, respectively. Similarly, if this system lies outdoor with operation frequency of 75 GHz and $L_t = L_r = 0.6$ meter, our optimized NULA deployment can maintain the EMG of 2 and 3 up to a communication distance of 461.5 meters and 106.9 meters, which are contrast to the respective distances of 161 meters and 61.9 meters for the ULA deployment. Some general observations can be made from Fig. \ref{Tau_vs_mu2and3overmu1_Fig}, as listed below.
    \begin{itemize}
        \item[$\bullet$] For any fixed $\Gamma$, the values of $\tau_{\min}^{(2)}$ and $\tau_{\min}^{(3)}$ for ULA-based systems increase with the numbers of antennas $M$ and $N$. This means that in a ULA-based mmWave LoS MIMO system with a given configuration (e.g., $D$, $L_t$ and $L_r$), increasing the system power gain by allocating more antennas will reduce the achievable EMG;
        \item[$\bullet$] For any fixed $\Gamma$, the values of $\tau_{\min}^{(2)}$ and $\tau_{\min}^{(3)}$ for optimized NULA-based systems remain constant when $M$ and $N$ increase. This property comes from a groupwise antenna deployment. It means that the newly added antennas can be completely utilized to provide power gain without affecting the achievable multiplexing gain;
        \item[$\bullet$] Given the same $M$ and $N$, the value of $\tau_{\min}^{(2)}$ (or $\tau_{\min}^{(3)}$) achieved by the optimized NULA is always smaller than that achieved by the conventional ULA, regardless the value of $\Gamma$. This indicates that the proposed optimized NULA is superior to the ULA in terms of achievable EMG, both asymptotically (when $\Gamma$ is sufficiently small, and consequently $\tau_{\min}^{(2)}$ or $\tau_{\min}^{(3)}$ will also be sufficiently small) and non-asymptotically (when $\Gamma$ takes finite values).
    \end{itemize}

\begin{figure}[t]
        \centering
        \scalebox{0.43}{\includegraphics{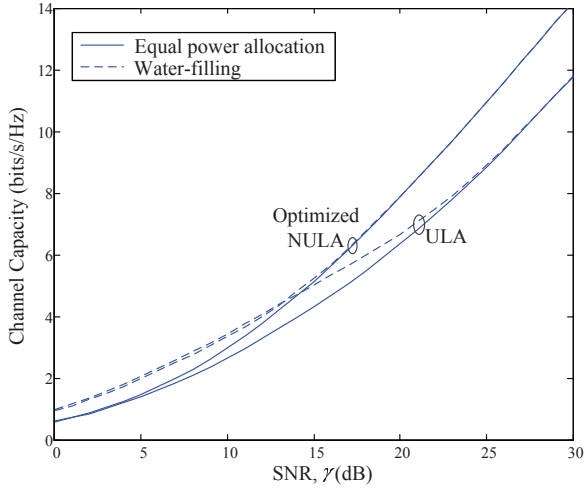}} \\
        \scalebox{0.75}{\includegraphics{a_K2.eps}} \\
        \scalebox{0.43}{\includegraphics{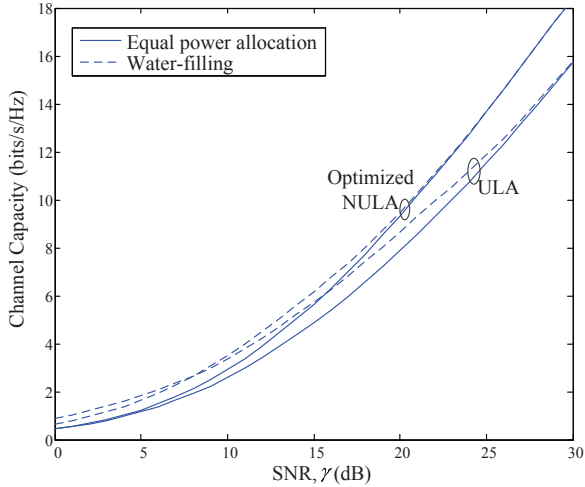}} \\
        \scalebox{0.75}{\includegraphics{b_K3.eps}} \\
        \caption{Capacity comparison between ULA- and optimized NULA-based mmWave LoS MIMO systems with $M \!=\! N \!=\! 24$. The communication distance is set such that $\tau = \tau_{\mathrm{min}}^{(2)}$ and $\tau = \tau_{\mathrm{min}}^{(3)}$, respectively in Figs. (a) and (b)}
        \label{Capacity_M24N24K2and3_Fig}
    \end{figure}

    The optimized NULA is also superior to ULAs in terms of capacities. In Fig. \ref{Capacity_M24N24K2and3_Fig} we plot the aforementioned system capacities achieved by both ULAs and the optimized NULAs\footnote{For practical reasons, we set a spacing of $\lambda/2 = 0.002$ meter, instead of $0$, between the adjacent antennas within each group. This adjustment only leads to marginal capacity loss.}  when the communication distance is set such that $\tau = \tau_{\mathrm{min}}^{(2)}$ and $\tau = \tau_{\mathrm{min}}^{(3)}$, respectively in Figs. \ref{Capacity_M24N24K2and3_Fig}(a) and \ref{Capacity_M24N24K2and3_Fig}(b). For convenience, we set $|\rho| \lambda/4 \pi D = 1/MN$, such that the total channel power is normalized, i.e., $tr(\bm{H}\bm{H}^H) = 1$. Thus the SNR $\gamma$ here represents the received SNR. We can clearly see the slope difference between the curves with ULAs and optimized NULAs for both the waterfilling capacity and that with equal power allocation (among the largest two or three eigenmodes), indicating that a higher effective multiplexing gain can be achieved using the proposed optimized NULAs, even in the non-asymptotic scenario.

\subsection{Groupwise PAT NULA Deployment}
\begin{figure}[t]
        \centering
        \scalebox{0.43}{\includegraphics{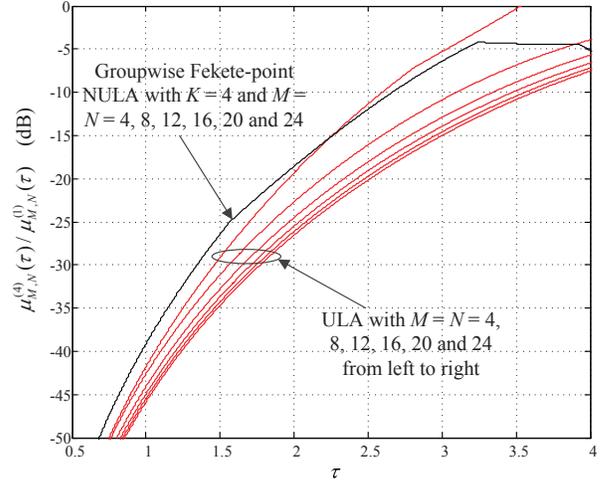}} \\
        \caption{The values of $\mu_{M,N}^{(4)}(\tau)/\mu_{M,N}^{(1)}(\tau)$ versus $\tau$ achieved by ULAs and optimized NULAs in mmWave LoS MIMO channels with variable numbers of transmit and receive antennas.}
        \label{Tau_vs_mu4overmu1_Fig}
    \end{figure}

    In this subsection, we discuss the possibility of achieving a higher effective multiplexing gain, i.e., $K \geq 4$. Fig. \ref{Tau_vs_mu4overmu1_Fig} plots the curves of $\mu_{M,N}^{(4)}(\tau)/\mu_{M,N}^{(1)}(\tau)$ versus $\tau$ achieved by ULAs and optimized NULAs in mmWave LoS MIMO channels with variable numbers of antennas. Similar to Fig. \ref{Tau_vs_mu2and3overmu1_Fig}, we can obtain from Fig. \ref{Tau_vs_mu4overmu1_Fig} the value of $\tau_{\min}^{(4)}$ for any given value of the threshold $\Gamma$ in each system. We can make the following observations from Fig. \ref{Tau_vs_mu4overmu1_Fig}.
    \begin{itemize}
        \item[$\bullet$] Given $\Gamma$, the value of $\tau_{\min}^{(4)}$ increases with the antenna numbers $M$ and $N$ in the ULA-based systems, but remains constant in the optmized NULA-based ones. This observation is the same as that made from Fig. \ref{Tau_vs_mu2and3overmu1_Fig} and reflects the superiority of the groupwise NULA deployment;
        \item[$\bullet$] For small values of $\Gamma$ (e.g., $\Gamma < -15$ dB), the value of $\tau_{\min}^{(4)}$ in the optimized NULA-based system is always smaller than that in the corresponding ULA-based system, indicating that the proposed groupwise Fekete-point NULA deployment is indeed asymptotically superior to the ULA one.
        \item[$\bullet$] When $\Gamma > -15$ dB, the value of $\tau_{\min}^{(4)}$ in the optimized NULA-based system becomes larger than its counterpart in the ULA-based system with $M = N = 4$, showing that the proposed NULA deployment in the previous sections is, though asymptotically optimal, suboptimal in some non-asymptotic scenarios.
    \end{itemize}

    \begin{figure}[t]
        \centering
        \scalebox{0.43}{\includegraphics{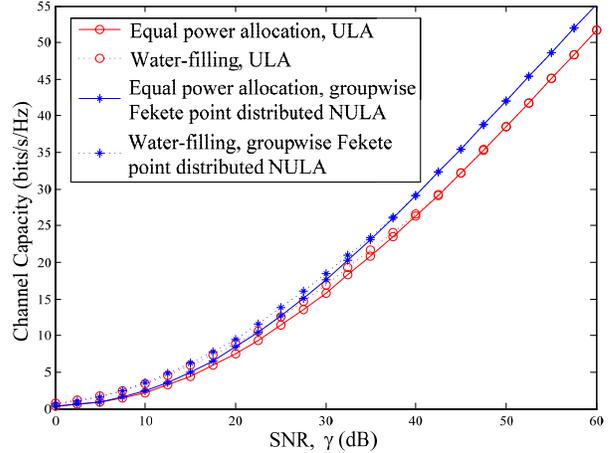}} \\
        \caption{Capacity comparison between ULA-based and optimized NULA-based mmWave LoS MIMO systems with $M = N = 24$ and $L_t = L_r = 0.6$ meter. The communication distance is set to be $D = 90.0686$ meters.}
        \label{Capacity_M24N24K4tao1d5696_Fig}
    \end{figure}

    The asymptotic optimality of the groupwise Fekete-point NULA deployment when $K = 4$ is also demonstrated in Fig. \ref{Capacity_M24N24K4tao1d5696_Fig} below in terms of capacity. Similar to Fig. \ref{Capacity_M24N24K2and3_Fig}, we consider a mmWave LoS MIMO system with $M = N = 24$, $L_t = L_r = 0.6$ meter and aim at achieving an EMG of $K = 4$. By setting the threshold at $\Gamma = -25$ dB, we can find from Fig. \ref{Tau_vs_mu4overmu1_Fig} that the corresponding value of $\tau_{\min}^{(4)}$ for the groupwise Fekete-point NULA deployment is $\tau_{\min}^{(4)} = 1.5696$, which corresponds to a communication distance of $D = 90.0686$ meters. Fig. \ref{Capacity_M24N24K4tao1d5696_Fig} plots the capacities of such a mmWave LoS MIMO system with $D = 90.0686$ meters that are achieved by both the groupwise Fekete-point NULAs and ULAs, from which we can easily see the superiority of the groupwise Fekete-point NULA deployment over the ULA one. Note that here we set a relatively small value for the threshold $\Gamma$ at $-25$ dB. This is because otherwise when $\Gamma > -15$ dB, the ULA deployment may outperform the groupwise Fekete-point NULA deployment in the mmWave system with $M = N = 4$ in terms of $\mu_{M,N}^{(4)}(\tau)/\mu_{M,N}^{(1)}(\tau)$, as seen from Fig. \ref{Tau_vs_mu4overmu1_Fig}. In addition, it is seen from Fig. \ref{Tau_vs_mu4overmu1_Fig} that, for the groupwise Fekete-point NULA deployment, the ratio of $\mu_{M,N}^{(4)}(\tau)/\mu_{M,N}^{(1)}(\tau)$ is always smaller than $-4.8$ dB. This indicates that, when we set $\Gamma > -4.8$ dB, we even cannot achieve the EMG of 4 using the groupwise Fekete-point NULA deployment practically. This problem is even more serious when the target EMG is higher than 4. Therefore, we need seek some more practical NULA design solutions, as detailed below.

    Recall from Section V-B that the Fekete-point distribution can be well approximated by the projected arch type (PAT) distribution with  angle $\theta = \theta_K$. Hence a straightforward option to the practical NULA design is to extend the proposed groupwise Fekete-point NULA deployment to the following \emph{groupwise PAT NULA deployment}: We still divide all the transmit/receive antenna into $K$ groups of approximately equal sizes with the minimum feasible antenna spacing in each group. Then, we require the centers of these groups to follow the PAT distribution with a certain angle $\theta$ and span the overall transmit/receive aperture. This groupwise PAT NULA deployment reduces to the groupwise Fekete-point NULA deployment when $\theta = \theta_K$. 
    Given the values of $K$ and $\Gamma$, we can easily find a proper value of $\theta$ via one-dimensional search to minimize $\tau_{\min}^{(K)}$.
    \begin{figure}[t]
        \centering
        \scalebox{0.43}{\includegraphics{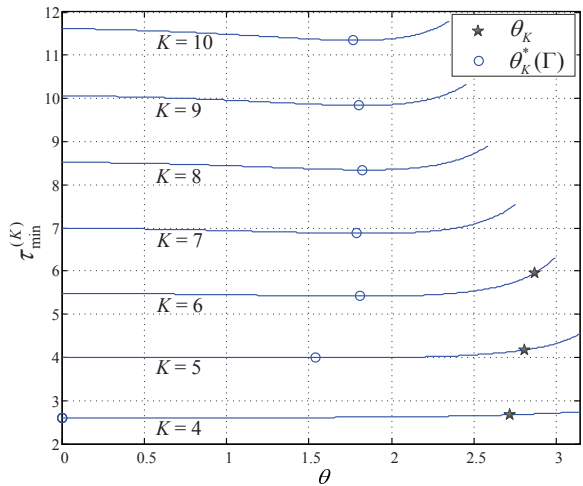}} \\
        \caption{The values of $\tau_{\min}^{(K)}$ achieved by the groupwise PAT NULA deployment with various angle $\theta$ and $\Gamma = -10$ dB.}
        \label{theta_vs_taumin_forPAT_Fig}
    \end{figure}

    Fig. \ref{theta_vs_taumin_forPAT_Fig} plots the values of $\tau_{\min}^{(K)}$ achieved by the general groupwise PAT NULA deployment with various angle $\theta$ when the threshold $\Gamma$ is set at a reasonable level of $\Gamma = -10$ dB. The optimal value of $\theta$ that minimizes $\tau_{\min}^{(K)}$, denoted by $\theta_{K}^{*}(\Gamma)$, is marked by ``$\bigcirc$'' in the figure. The values of $\{\theta_K\}$ that correspond to the groupwise Fekete-point NULA deployment are also marked in Fig. \ref{theta_vs_taumin_forPAT_Fig} by ``$\bigstar$''. Note that with $\Gamma = -10$ dB, $\tau_{\min}^{(K)}$ does not exist for $K \geq 7$ in systems with the groupwise Fekete-point NULA deployment, and so the corresponding $\{\theta_K\}$ are not shown in the figure. From Fig. \ref{theta_vs_taumin_forPAT_Fig} we can see that when $K \geq 5$, we should set a non-zero value of $\theta$ for the general groupwise PAT NULA deployment to achieve the minimum $\tau_{\min}^{(K)}$. 
    It is also seen from Fig. \ref{theta_vs_taumin_forPAT_Fig} that the value of $\tau_{\min}^{(K)}$ is almost unchanged when $\theta$ varies around its optimal value, i.e., $\theta_{K}^{*}(\Gamma)$, indicating that the proposed groupwise PAT NULA deployment is robust to calibration errors in real systems.


\section{Extension to Non-Uniform Rectangular Antenna Arrays}
    It is worth noting that, following a similar derivation as that in Section VII of \cite{Peng_TWC14}, we can readily extend the above discussion to the system with two-dimensional non-uniform rectangular antenna arrays (NURAs) at both the transmitter and receiver, where the rows (columns) of the transmit and receive NURAs are parallel and aligned with each other. In this case, the system EMG can be enhanced by letting each row/column of the transmit/receive NULAs to follow a groupwise PAT deployment. The detailed discussions are omitted here due to space limitation.

\section{Conclusions}
    In this paper, we investigated the NULA deployment optimization in mmWave LoS MIMO channels for maximizing the system EMG. Our analysis shows that the highest multiplexing gain can be achieved by a groupwise Fekete-point NULA deployment in the limit when the transmit-receive distance is very large relative to the aperture sizes of the transmit/receive NULAs. We also developed a simple and accurate approximation for the Fekete-point distribution using the PAT distribution, which can be characterized by a single angle parameter. Finally, we discussed the NULA deployment in some practical mmWave communication scenarios. We numerically developed a more general array deployment, referred to as groupwise PAT NULA deployment. Numerical results are provided to demonstrate the performance gain of the optimized NULA deployments over the conventional ULAs. The results in this paper provide useful insights into the design of future wireless communications systems operating at tens-of-Gigabits/second data rates. Such high-speed systems may find use in a wide spectrum of applications, including wireless backhaul, last-mile access, network recovery, campus LAN and storage access.
%


%

\section{Appendices}
\subsection{Proof of Theorem 1}
    To prove (\ref{Eigvalue_slope}), let us first return to (\ref{h_hat}) and perform Taylor expansion on $\hat{h}_{m,n}$, i.e.,
    \begin{equation}
        \hat{h}_{m,n} = e^{j\tau \alpha_{r,m}\alpha_{t,n}} = \sum\nolimits_{i=0}^{\infty}\frac{\left(j\tau\alpha_{r,m}\alpha_{t,n}\right)^i}{i!}.
        \label{h_hat_Taylor}
    \end{equation}
    From (\ref{h_hat_Taylor}), we can decompose $\hat{\bm{H}}$ as
    \begin{equation}
        \hat{\bm{H}} = \bm{A}\bm{T}\bm{B}^T
        \label{H_hat_Decomposition1}
    \end{equation}
    where $\bm{A} = \bm{C}^{(r)}_{M\times \infty}$, $\bm{B} = \bm{C}^{(t)}_{N \times \infty}$ (see 
    (\ref{QR_decomp_Tx}) for definitions) and $\bm{T} = \text{diag}\{t_1,t_2,\cdots\}$ is an $\infty$-by-$\infty$ diagonal complex matrix with $t_m = (j\tau)^{m-1}/(m-1)!$.

    For the ease of derivation below, we further divide matrices $\bm{A}$, $\bm{B}$ and $\bm{T}$ into sub-matrices as, respectively, $\bm{A} = (\bm{A}_1 \bm{A}_2, \cdots)$, $\bm{B} = (\bm{B}_1 \bm{B}_2, \cdots)$ and $\bm{T} = \text{diag}(\bm{T}_1, \bm{T}_2, \cdots)$ where $\bm{A}_m$ is the $m$-th $M$-by-$M$ sub-matrix of $\bm{A}$, $\bm{B}_m$ the $m$-th $N$-by-$M$ sub-matrix of $\bm{B}$ and $\bm{T}_m$ the $m$-th $M$-by-$M$ diagonal sub-matrix of $\bm{T}$. Define $\hat{\bm{H}}_m \doteq \bm{A}_m\bm{T}_m\bm{B}_m^T$. Then (\ref{H_hat_Decomposition1}) can be rewritten as
    \begin{equation}
        \hat{\bm{H}} = \sum\nolimits_{m=1}^{\infty} \bm{A}_m\bm{T}_m\bm{B}_m^T = \sum\nolimits_{m=1}^{\infty} \hat{\bm{H}}_m.
        \label{H_decomp}
    \end{equation}

    Next, we focus on the singular values of $\hat{\bm{H}}_1$, or equivalently the eigenvalues of $\hat{\bm{G}} \doteq \hat{\bm{H}}_1\hat{\bm{H}}_1^H$. Denote by $\hat{\mu}_{M,N}^{(m)}(\tau)$ the $m$-th largest eigenvalue of $\hat{\bm{G}}$. We have
    \begin{eqnarray}
        \nonumber
        & \ & \lim_{\tau \to 0} \frac{\text{ln}\left(\prod_{m=1}^{M}\hat{\mu}_{M,N}^{(m)}(\tau)\right)}{\text{ln}\tau} = \lim_{\tau \to 0} \frac{\text{ln}\left(\text{det}(\hat{\bm{G}})\right)}{\text{ln}\tau}\\
        \nonumber
        & = & \lim_{\tau \to 0} \frac{\text{ln}\left(\text{det}(\bm{A}_1\bm{T}_1\bm{B}_1^T\bm{B}_1\bm{T}_1^H\bm{A}_1^T)\right)}{\text{ln}\tau} \\
        \nonumber
        & = & \lim_{\tau \to 0} \frac{\text{ln}\left(\text{det}(\bm{A}_1^T\bm{A}_1\bm{T}_1\bm{B}_1^T\bm{B}_1\bm{T}_1^H)\right)}{\text{ln}\tau} \\
        \nonumber
        & = & \lim_{\tau \to 0} \frac{\text{ln}\left(\text{det}(\bm{A}_1^T\bm{A}_1)\text{det}(\bm{T}_1)\text{det}(\bm{B}_1^T\bm{B}_1)\text{det}(\bm{T}_1^H) \right)}{\text{ln}\tau} \\
        \nonumber
        & = & \lim_{\tau \to 0} \frac{\text{ln}\left(\text{det}(\bm{A}_1^T\bm{A}_1)\text{det}(\bm{B}_1^T\bm{B}_1)\frac{\tau^{M(M-1)}}{\big(\prod_{m=1}^{M-1}m!\big)^{2}} \right)}{\text{ln}\tau} \\
        & = & M(M-1).
        \label{EigProd_Slope}
    \end{eqnarray}
    On the other hand, recalling the fact $\bm{A}_1 = \bm{C}^{(r)}_{M \times M}$ and the QR decomposition 
    (\ref{QR_decomp_Tx}), we define
    \begin{equation}
        \widetilde{\bm{G}} \!\!\doteq\!\! (\bm{Q}^{(r)}_M)^T\hat{\bm{G}}\bm{Q}^{(r)}_M \!\!=\!\! \bm{R}^{(r)}_{M \times M}\bm{T}_1\bm{B}_1^T\bm{B}_1\bm{T}_1^H(\bm{R}^{(r)}_{M \times M})^T.
    \end{equation}
    Then $\widetilde{\bm{G}}$ has the same eigenvalues as $\hat{\bm{G}}$ (i.e., $\{\hat{\mu}_{M,N}^{(m)}(\tau), m = 1,2,\cdots, M\}$). In addition, since $\bm{R}^{(r)}_{M \times M}$ is upper-triangular, the $m$-th diagonal entry of $\widetilde{\bm{G}}$, denoted by $\widetilde{g}_m$, satisfies,
    \begin{equation}
        \lim_{\tau \to 0} \frac{\text{ln}\widetilde{g}_m}{\text{ln}\tau}  = 2(m-1), \forall m = 1, 2, \cdots, M.
        \label{G_dia_entry}
    \end{equation}
    From (\ref{G_dia_entry}), we can upper-bound the asymptotic slope of $\hat{\mu}_{M,N}^{(m)}(\tau)$ by
    \begin{eqnarray}
        \nonumber
        & \ & \lim_{\tau \to 0} \frac{\text{ln}\left(\hat{\mu}_{M,N}^{(m)}(\tau)\right)}{\text{ln}\tau}
        \leq \lim_{\tau \to 0} \frac{\text{ln}\left(\sum_{l=m}^{M}\hat{\mu}_{M,N}^{(l)}(\tau)\right)}{\text{ln}\tau} \\
        & \!\!\!\! \leq \!\!\!\! & \lim_{\tau \to 0} \frac{\text{ln}\left(\sum_{l=m}^{M}\widetilde{g}_l)\right)}{\text{ln}\tau} = 2(m-1)
        \label{Eigvalue_LB}
    \end{eqnarray}
    where the second inequality follows Theorem 4.3.26 of \cite{MatrixBook} (pp.195). Based on (\ref{Eigvalue_LB}), we obtain
    \begin{eqnarray}
        \nonumber
        & \ & \lim_{\tau \to 0}\frac{\text{ln}\left(\prod\limits_{m-1}^{M}\hat{\mu}_{M,N}^{(m)}(\tau)\right)}{\text{ln}\tau}
        = \lim_{\tau \to 0}\sum_{m-1}^{M}\frac{\text{ln}\left(\hat{\mu}_{M,N}^{(m)}(\tau)\right)}{\text{ln}\tau} \\
        &\!\! \leq \!\!&  \sum_{m-1}^{M}2(m-1) = M(M-1).
         \label{EigProd_LB}
    \end{eqnarray}
    Note that (\ref{EigProd_LB}) holds with equality if and only if all the inequalities in (\ref{Eigvalue_LB}) holds with equality for $\forall m = 1, 2,\cdots,M$. In fact, the equality in (\ref{EigProd_LB}) does hold due to (\ref{EigProd_Slope}), yielding
    \begin{equation}
        \lim_{\tau \to 0} \frac{\text{ln}\left(\hat{\mu}_{M,N}^{(m)}(\tau)\right)}{\text{ln}\tau} = 2(m-1), \forall m = 1,2,\cdots, M.
    \end{equation}

    Finally, we need to check the relationship between $\hat{\mu}_{M,N}^{(m)}(\tau)$ and $\mu_{M,N}^{(m)}(\tau)$. Return to (\ref{H_decomp}). Since $\hat{\bm{H}}$ can be viewed as a perturbation form of $\hat{\bm{H}}_1$, i.e., $\hat{\bm{H}} = \hat{\bm{H}}_1 + \sum\nolimits_{m=2}^{\infty} \hat{\bm{H}}_m$, according to Weyl's Perturbation Theorem \cite{Wely_1912}\cite{Pertubation90}, we have
    \begin{equation}
        \left|\sqrt{\mu_{M,N}^{(m)}(\tau)} - \sqrt{\hat{\mu}_{M,N}^{(m)}(\tau)} \right| \leq   \|\sum\nolimits_{m=2}^{\infty} \hat{\bm{H}}_m\|.
        \label{Weyl_Theorem}
    \end{equation}
    Recall the definition of $\hat{\bm{H}}_m$. We have
    \begin{eqnarray}
        \nonumber
        & \    &\lim_{\tau \to 0} \frac{\|\sum\nolimits_{m=2}^{\infty} \hat{\bm{H}}_m\|}{\tau^{M-1}} \\
        \nonumber
        & \leq & \lim_{\tau \to 0} \frac{\sum\nolimits_{m=2}^{\infty}\|\bm{A}_m\|\!\cdot\!\|\bm{T}_m\|\!\cdot\!\|\bm{B}_m^T\|}{\tau^{M-1}} \\
        & =    & \lim_{\tau \to 0} \frac{\sum\nolimits_{m=2}^{\infty}\|\bm{A}_m\|\!\cdot\!\frac{\tau^{M(m-1)}}{(M(m-1))!}\!\cdot\!\|\bm{B}_m^T\|}{\tau^{M-1}} = 0.
    \end{eqnarray}
    Combining this and (\ref{Weyl_Theorem}), we obtain
    \begin{eqnarray}
        \nonumber
        \ & \ & \lim_{\tau \to 0} \left|\sqrt{\mu_{M,N}^{(m)}\!(\tau)/\tau^{2(m-1)}} \!\!-\!\! \sqrt{\hat{\mu}_{M,N}^{(m)}\!(\tau)/\tau^{2(m-1)}}\right| \\
        \ & \leq & \lim_{\tau \to 0} \frac{\sum_{m=2}^{\infty}\| \hat{\bm{H}}_m\|}{\tau^{M-1}} = 0,
    \end{eqnarray}
    which indicates that the difference between $\mu_{M,N}^{(m)}$ and $\hat{\mu}_{M,N}^{(m)}$ is a higher-order infinitesimal term of $\tau^{2(m-1)}$. Hence we have
    \begin{eqnarray}
        \nonumber
        \lim_{\tau \to 0} \frac{\text{ln}\left(\mu_{M,N}^{(m)}(\tau)\right)}{\text{ln}\tau}
        & \!\!\!\! = \!\!\!\! &
        \lim_{\tau \to 0} \frac{\text{ln}\left(\hat{\mu}_{M,N}^{(m)}(\tau)\right)}{\text{ln}\tau} \\
        & \!\!\!\! = \!\!\!\! & 2(m\!-\!1), \ \forall m \!=\! 1, 2, \cdots, M,
    \end{eqnarray}
    which completes the proof.


\subsection{Proof of Theorem 2}
    We first introduce some notations that are useful in the proof below. Let $\bm{u}_{M}^{(m)}(\tau)$ be the eigenvector of $\bm{G}_{M,N}(\tau)$ corresponding to $\mu_{M,N}^{(m)}(\tau)$. Define
    \begin{equation}
        \dot{\bm{u}}_M^{(m)} \doteq \lim_{\tau \to 0} \bm{u}_M^{(m)}(\tau).
    \end{equation}
    Then $\{\dot{\bm{u}}_M^{(m)}| m = 1, 2, \cdots, M\}$ form a basis for the $M$ dimensional space, and the $m$-th column of $\bm{Q}^{(r)}_M$, denoted by $\bm{q}_M^{(m)}$ can be represented as
    \begin{equation}
        \bm{q}_M^{(m)} = \sum\nolimits_{n=1}^{M}\varepsilon_{m,n}\dot{\bm{u}}_M^{(n)}, \forall m = 1, 2, \cdots, M
    \end{equation}
    where $\{\varepsilon_{m,n}\}$ are real numbers satisfying $\sum_{n=1}^{M} \varepsilon_{m,n}^2 = 1, \forall m = 1, 2, \cdots, M$.

    Theorem 2 can be proved by contradiction starting from $m = M$. Assume that $\bm{u}_M^{(M)}(\tau)$ does not converge to $\bm{q}_M^{(M)}$ as $\tau \to 0$, i.e., $\bm{q}_M^{(M)} \neq \dot{\bm{u}}_M^{(M)}$. Then we have $\varepsilon_{M,M} \neq 1$ and $\exists n^* \ (n^* < M)$ such that $\varepsilon_{M,n^*} \neq 0$. This leads to
    \begin{eqnarray}
        \nonumber
        & \ & \lim_{\tau \to 0} \widetilde{g}_M = \lim_{\tau \to 0} \big(\bm{q}_M^{(M)}\big)^T \bm{G}_{M,N}(\tau) \bm{q}_M^{(M)} \\
        \nonumber
        & = & \lim_{\tau \to 0} \left(\sum_{n=1}^{M}\varepsilon_{M,n}\dot{\bm{u}}_M^{(n)}\right)^T \bm{G}_{M,N}(\tau) \left(\sum_{n=1}^{M}\varepsilon_{M,n}\dot{\bm{u}}_M^{(n)}\right) \\
        \nonumber
        & = & \lim_{\tau \to 0} \sum_{n=1}^{M}\varepsilon_{M,n}^2\big(\dot{\bm{u}}_M^{(n)}\big)^T \bm{G}_{M,N}(\tau) \dot{\bm{u}}_M^{(n)} \\
        \nonumber
        & = & \lim_{\tau \to 0} \sum_{n=1}^{M}\varepsilon_{M,n}^2 \mu_{M,N}^{(n)}(\tau) \geq \varepsilon_{M,n^*}^2 \mu_{M,N}^{(n^*)}(\tau),
    \end{eqnarray}
    which indicates that
    \begin{eqnarray}
        \nonumber
        \lim_{\tau \to 0}\frac{\text{ln}\widetilde{g}_M}{\text{ln}\tau}
        & \leq & \lim_{\tau \to 0}\frac{\text{ln}\big(\varepsilon_{M,n^*}^2 \mu_{M,N}^{(n^*)}(\tau)\big)}{\text{ln}\tau} \\
        & = & 2(n^*-1) < 2(M-1).
        \label{AsyEigVector_Proof1}
    \end{eqnarray}
    Clearly, (\ref{AsyEigVector_Proof1}) is inconsistent with (\ref{G_dia_entry}). Hence we conclude that $\varepsilon_{M,M} = 1$ and $\varepsilon_{M,n} = 0, \forall n \neq M$, leading to $\dot{\bm{u}}_M^{(M)} = \bm{q}_M^{(M)}$.

    Next, we consider $m = M-1$. From the facts $\dot{\bm{u}}_M^{(M)} = \bm{q}_M^{(M)}$ and $\dot{\bm{u}}_M^{M-1} \perp \dot{\bm{u}}_M^{(M)}$, we have $\varepsilon_{M-1,M} = 0$. Similar to the above argument for the case of $m = M$, we can conclude that the assumption $\dot{\bm{u}}_M^{(M-1)} \neq \bm{q}_M^{(M-1)}$ will lead to
    \begin{equation}
        \lim_{\tau \to 0} \frac{\text{ln}\widetilde{g}_{M-1}}{\text{ln}\tau} > 2(M -2),
    \end{equation}
    which is again inconsistent with (\ref{G_dia_entry}). By adopting this arguement recursively, we can show that $\dot{\bm{u}}_M^{(m)} = \bm{q}_M^{(m)}$ for all $m = 1, 2, \cdots, M$. Hence the proof is completed.

\subsection{Proof of Corollary 1}
    The from of (\ref{EigValueApprox}) directly follows from Theorem 1 and here we only prove the coefficient. Since $\bm{G}_{M,N}(\tau) = \bm{\hat{H}}\bm{\hat{H}}^H$ from (\ref{G_matrix}), Theorem 2 implies that the left singular vectors of $\bm{\hat{H}}$ converge to the columns of $\bm{Q}^{(r)}_M$ (apart from a phase factor) when $\tau \to 0$. Similarly, we can also show that, as $\tau \to 0$, the right singular vectors of $\bm{\hat{H}}$ converge to the first $M$ columns of $\bm{Q}^{(t)}_N$ (also apart from a phase factor). Denote by $\bm{p}^{(m)}_N$ the $m$-th column of unitary matrix $\bm{Q}^{(t)}_N$. Then combining this and (\ref{H_hat_Decomposition1}), we obtain
    \begin{eqnarray}
        \nonumber
        \lim_{\tau \to 0} \frac{\mu_M^{(m)}(\tau,N)}{\tau^{2(m-1)}}
        & \!\!\!\!=\!\!\!\! & \lim_{\tau \to 0} \frac{\Big| \big(\bm{q}_M^{(m)}\big)^T \hat{\bm{H}}\bm{p}_N^{(m)}\Big|^2}{\tau^{2(m-1)}} \\
        & \!\!\!\!=\!\!\!\! & \lim_{\tau \to 0} \frac{\Big| \big(\bm{q}_M^{(m)}\big)^T \bm{A}\bm{T}\bm{B}^T\bm{p}_N^{(m)}\Big|^2}{\tau^{2(m-1)}}.
    \label{AsyEigCoefficient1}
    \end{eqnarray}

    Similar to (\ref{QR_decomp_Tx}), 
    we perform QR decomposition on $\bm{A}$ and $\bm{B}$ as
    \begin{equation}
        \bm{A} = \bm{C}^{(r)}_{M \times \infty} = \bm{Q}^{(r)}_M \bm{R}^{(r)}_{M \times \infty}
        \ \ \mathrm{and} \ \
        \bm{B} = \bm{C}^{(t)}_{N \times \infty} = \bm{Q}^{(t)}_N \bm{R}^{(t)}_{N \times \infty}
        \label{C_QRDecompGeneral_Tx}
    \end{equation}
    respectively. It is worth to point out that, here we use the same notations of $\bm{Q}^{(r)}_M$ and $\bm{Q}^{(t)}_N$ as those in (\ref{QR_decomp_Tx}). 
    This can be interpreted as follows by taking $\bm{Q}^{(r)}_M$ as an example. From the operation of the QR decomposition, the unitary matrix $\bm{Q}^{(r)}_M$ in 
    (\ref{C_QRDecompGeneral_Tx}) only depends on the first $M$ columns of $\bm{C}^{(r)}_{M \times \infty}$, which are the same as those of $\bm{C}^{(r)}_{M \times M}$ in (\ref{QR_decomp_Tx}). 
     Similarly, the first $M$ columns of $\bm{R}^{(r)}_{M \times \infty}$ in 
    (\ref{C_QRDecompGeneral_Tx}) are also the same at those of $\bm{R}^{(r)}_{M \times M}$ in (\ref{QR_decomp_Tx}). 
    (The same relationship holds between $\bm{R}^{(t)}_{N \times \infty}$ and $\bm{R}^{(t)}_{N \times N}$.) Denote by $r^{(t)}_{m,n}$ and $r^{(r)}_{m,n}$, respectively, the $(m,n)$-th elements of matrices $\bm{R}^{(t)}_{N \times \infty}$ and $\bm{R}^{(r)}_{M \times \infty}$. We can rewrite (\ref{AsyEigCoefficient1}) based on 
    (\ref{C_QRDecompGeneral_Tx}) as
    \begin{eqnarray}
        \nonumber
        & \ &  \lim_{\tau \to 0} \frac{\mu_{M,N}^{(m)}(\tau)}{\tau^{2(m-1)}} \\
        \nonumber
        & = & \lim_{\tau \to 0} \left|\frac{\big(\bm{q}_M^{(m)}\big)^T \bm{Q}^{(r)}_M\bm{R}^{(r)}_{M \times \infty}\bm{T}(\bm{R}^{(t)}_{N \times \infty})^T(\bm{Q}^{(t)}_N)^T\bm{p}_N^{(m)}}{\tau^{m-1}}\right|^2 \\
        \nonumber
        & = & \lim_{\tau \to 0} \left|\frac{\sum_{n=m}^{\infty}r^{(r)}_{m,n} \cdot \frac{(j\tau)^{m-1}}{(m-1)!} \cdot r^{(t)}_{m,n}}{\tau^{m-1}}\right|^2 \\
        & = & \left( \frac{r^{(r)}_{m,m}r^{(t)}_{m,m}}{(m-1)!} \right)^2.
        \label{AsyEigCoefficient2}
    \end{eqnarray}
    Finally, by redefining $r^{(r)}_{m,m} \doteq r^{(r)}_m$ and $r^{(t)}_{m,m} \doteq r^{(t)}_m$ in (\ref{AsyEigCoefficient2}) for simplicity, we obtain (\ref{EigValueApprox}). 

\subsection{Proof of Theorem 3}
    The proof of (\ref{R_diag_expression1}) is apparent. From the operation of QR decomposition, $r^{(r)}_1$ equals the norm of the first column in $\bm{C}^{(r)}_{M \times M}$, i.e., a length-$M$ all-one vector. Thus (\ref{R_diag_expression1}) holds.

    Now we consider the proof of (\ref{R_diag_expression2}). From 
    (\ref{QR_decomp_Tx}), we have
    \begin{eqnarray}
        \nonumber
        \text{det}\left(\big(\bm{C}^{(r)}_{M \times k}\big)^T\!\bm{C}^{(r)}_{M \times k}\right)
        &\!\!\!\!\!\! = \!\!\!\!\!\!&  \text{det}\left(\big(\bm{R}^{(r)}_{M \times k}\big)^T\!\big(\bm{Q}^{(r)}_M\big)^T\!\bm{Q}^{(r)}_M\bm{R}^{(r)}_{M \times k}\right) \\
        \nonumber
        &\!\!\!\!\!\! = \!\!\!\!\!\!& \text{det}\left(\big(\bm{R}^{(r)}_{M \times k}\big)^T\!\bm{R}^{(r)}_{M \times k}\right)
         = \prod_{i = 1}^{k}\!\big(r^{(r)}_i\big)^2,
    \end{eqnarray}
    which indicates that
    \begin{eqnarray}
        \nonumber
        r^{(r)}_k & \!\!\!=\!\!\! & \left(\frac{\prod_{i=1}^{k}\!\big(r^{(r)}_i\big)^2}{\prod_{i=1}^{k-1}\!\big(r^{(r)}_i\big)^2}\right)^{1/2}  \\
        & \!\!\!=\!\!\! & \left(\frac{\text{det}\left(\big(\bm{C}^{(r)}_{M \times k}\big)^T\!\bm{C}^{(r)}_{M \times k}\right)}{\text{det}\left(\big(\bm{C}^{(r)}_{M \times (k-1)}\big)^T\!\bm{C}^{(r)}_{M \times (k-1)}\right)}\right)^{1/2}.
        \label{R_diag_expression3}
    \end{eqnarray}
    Meanwhile, from the Cauchy-Binet formula \cite{CauchyBient}, we have
    \begin{eqnarray}
        \nonumber
        & \!\!\!\!\!\!\ \!\!\!\!\!\! & \text{det}\left(\big(\bm{C}^{(r)}_{M \times K}\big)^T\!\bm{C}^{(r)}_{M \times K}\right) \\
        & \!\!\!\!\!\!= \!\!\!\!\!\! & \sum\limits_{\overset{\mathcal{S} \subset \{1, 2, \cdots, M\},}{|\mathcal{S}| = K}}\text{det}\left(\big(\bm{C}^{(r)}_{M \times K}(\mathcal{S})\big)^T\right)\cdot\text{det}\left(\bm{C}^{(r)}_{M \times K}(\mathcal{S})\right)
        \label{CauchyBientFormula}
    \end{eqnarray}
    where $\bm{C}^{(r)}_{M \times K}(\mathcal{S})$ is a $K$-by-$K$ sub-matrix of $\bm{C}^{(r)}_{M \times K}$ containing $K$ rows of the latter indicated by the set $\mathcal{S}$. Clearly, $\bm{C}^{(r)}_{M \times K}(\mathcal{S})$ is also a Vandermonde matrix, whose determinant is given by
    \begin{eqnarray}
        \nonumber
        \text{det}\left(\bm{C}^{(r)}_{M \times K}(\mathcal{S})\right) & = & \text{det}\left(\big(\bm{C}^{(r)}_{M \times K}(\mathcal{S})\big)^T\right) \\
        & = & \prod\limits_{\overset{i<j,}{i,j\in \mathcal{S}}}\big(\alpha_{r,j} - \alpha_{r,i}\big).
        \label{VandMatrixDet}
    \end{eqnarray}
    Substituting (\ref{VandMatrixDet}) into (\ref{CauchyBientFormula}) and (\ref{R_diag_expression3}) in turn, we can obtain (\ref{R_diag_expression2}). This completes the proof.

\subsection{Proof of Theorem 4}
    Before proving Theorem 4, we first introduce a good property of the Fekete points in polynomial interpolation. Recall that $\{\gamma_{K,k}\}$ are a set of $K$ Fekete points that maximize a $K$-dimensional Vandermonde matrix over $[-1, 1]$. Define
    \begin{equation}
        l_{K,k}(x) = \frac{\prod\nolimits_{i \neq k} (x - \gamma_{K,i})}{\prod\nolimits_{i \neq k}(\gamma_{K,k} - \gamma_{K,i})}
        \label{Lag_interp_poly}
    \end{equation}
    as the associated fundamental (or cardinal) Lagrange interpolating polynomial. Then for any polynomial function $f(x)$ with degree less than $K$, we always have \cite{Bos_Math01}
    \begin{equation}
        f(x) = \sum\nolimits_{k=1}^{K}l_k(x)f(\gamma_{K,k}).
        \label{Lag_Interpolation}
    \end{equation}

    Now let us return to Theorem 4. From the proof of Theorem 3, it is easy to see that the target function in \textbf{P5} can be written into a matrix form as
    \begin{equation}
        \sum\nolimits_{\overset{\mathcal{S} \subset \{1, 2, \cdots, M\},}{|\mathcal{S}| = K}}\prod\nolimits_{\overset{i<j,}{i,j\in \mathcal{S}}}\big(\alpha_j - \alpha_i\big)^2 =\text{det}\left(\big(\bm{C}_{M \times K}^{(r)}\big)^T\bm{C}_{M \times K}^{(r)}\right).
        \label{P3_Matrixform}
    \end{equation}
    Using (\ref{Lag_Interpolation}), we represent each entry in $\bm{C}_{M \times K}^{(r)}$, i.e., $\alpha_m^i$, as
    \begin{equation}
        \alpha_m^i = \sum\nolimits_{k=1}^K l_{K,k}(\alpha_m) \gamma_{K,k}^i.
    \end{equation}
    Thus we have
    \begin{equation}
        \bm{C}_{M \times K}^{(r)} = \bm{L}_{M \times K} \bm{\Gamma}_{K \times K}
    \end{equation}
    where the entry at the $m$-th row and $k$-th column of $\bm{L}_{M \times K}$ is $l_{K,k}(\alpha_m)$ and $\bm{\Gamma}_{K \times K}$ is a $K \times K$ Vandermonde matrix generated by $\bm{\gamma}_K = (\gamma_{K,1}, \gamma_{K,2, \cdots, \gamma_{K,K}})$. Substituting this into (\ref{P3_Matrixform}), we obtain
    \begin{eqnarray}
        \nonumber
        & \ & \text{det}\left(\big(\bm{C}_{M \!\times\! K}^{(r)}\big)^T\!\bm{C}_{M \!\times\! K}^{(r)}\right) \\
        \nonumber
        & = & \text{det}\left(\big(\bm{L}_{M \times K} \bm{\Gamma}_{K \times K}\big)^T\bm{L}_{M \times K} \bm{\Gamma}_{K \times K}\right) \\
        \nonumber
        & = & \text{det}\left(\bm{\Gamma}_{K \times K}^T\bm{L}_{M \times K}^T\bm{L}_{M \times K} \bm{\Gamma}_{K \times K}\right) \\
        \nonumber
        & = & \text{det}\left(\bm{\Gamma}_{K \times K}^T\right)\cdot\text{det}\left(\bm{L}_{M \times K}^T\bm{L}_{M \times K} \right) \cdot \text{det}\left(\bm{\Gamma}_{K \times K}\right) \\
        \nonumber
        & = & f_{K,K}^2(\bm{\gamma}_K)\cdot \text{det}\left(\bm{L}_{M \times K}^T\bm{L}_{M \times K} \right) \\
        \nonumber
        & \leq & f_{K,K}^2(\bm{\gamma}_K)\prod_{k=1}^K \left(\sum_{m=1}^M l_{K,k}^2(\alpha_m)\right) \\
        & \leq & f_{K,K}^2(\bm{\gamma}_K)\left(\frac{1}{K}\sum_{k=1}^K \sum_{m=1}^M l_{K,k}^2(\alpha_m)\right)^K.
        \label{Determinant_UB}
    \end{eqnarray}
    In \cite{Bos_Math01}, it is shown that the polynomials $\{l_{K,k}(x)\}$ defined in (\ref{Lag_interp_poly}) always satisfy
    \begin{equation}
        \sum\nolimits_{k=1}^K l_{K,k}(x)^2 \leq 1 \ \ \ \forall x \in [-1, 1]
    \end{equation}
    where the equality holds only when $x = \gamma_{K,k}, \forall k$. Substituting this into (\ref{Determinant_UB}), we further upper bound the latter as
    \begin{eqnarray}
        \nonumber
        & \ & \text{det}\left(\big(\bm{C}_{M \!\times\! K}^{(r)}\big)^T\!\bm{C}_{M \!\times\! K}^{(r)}\right) \\
        \nonumber
        & \leq & f_{K,K}^2(\bm{\gamma}_K)\left(\frac{1}{K}\sum_{m=1}^M \left(\sum_{k=1}^K l_{K,k}^2(\alpha_m)\right)\right)^K \\
        & \leq & f_{K,K}^2(\bm{\gamma}_K)\left(\frac{M}{K}\right)^K.
        \label{Determinant_UB2}
    \end{eqnarray}

    Now we set the values of $\{\alpha_m|m=1,2,\cdots, M\}$ according to (\ref{Opt_alpha}) when $K$ divides $M$. Then it is easy to verify that both the two inequalities in (\ref{Determinant_UB2}) hold simultaneously, indicating that the setting of $\{\alpha_m|m=1,2,\cdots, M\}$ in Theorem 4 is optimal. This completes the proof.

\ifCLASSOPTIONcaptionsoff
  \newpage
\fi



%

%


\begin{IEEEbiography}[{\includegraphics[width=1in,height=1.25in,clip,keepaspectratio]{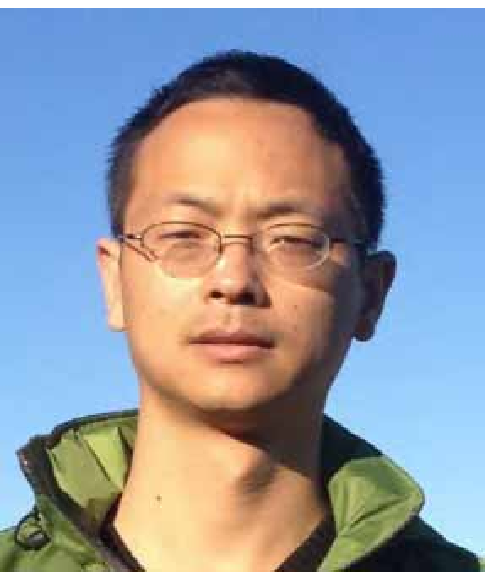}}]{Peng Wang} (S'05-M'10)
received the B. Eng. degree in telecommunication engineering and M. Eng. degree in information engineering, from Xidian University, Xi'an, China, in 2001 and 2004, respectively, and the Ph.D. degree in electronic engineering from the City University of Hong Kong, Hong Kong SAR, in 2010.

He was a Research Fellow with the City University of Hong Kong and a visiting Post-Doctor Research Fellow with the Chinese University of Hong Kong, Hong Kong SAR, both from 2010 to 2012, and a Research Fellow with the Center of Excellence in Telecommunications, School of Electrical and Information Engineering, the University of Sydney, Australia, from 2012 to 2015. Since 2015, he has been with Huawei Technologies, Sweden AB, where he is currently a Senior Research Engineer. He has published over 50 peer-reviewed research papers in the leading international journals and conferences. His research interests include channel and network coding, information theory, iterative multi-user detection, MIMO techniques and millimetre-wave communications.

Dr. Wang won the Best Paper Award at the 2014 IEEE International Conference on Communications (ICC). He has served on a number of technical programs for international conferences such as ICC and the Wireless Communications and Networking Conference (WCNC).
\end{IEEEbiography}

\begin{IEEEbiography}[{\includegraphics[width=1in,height=1.25in,clip,keepaspectratio]{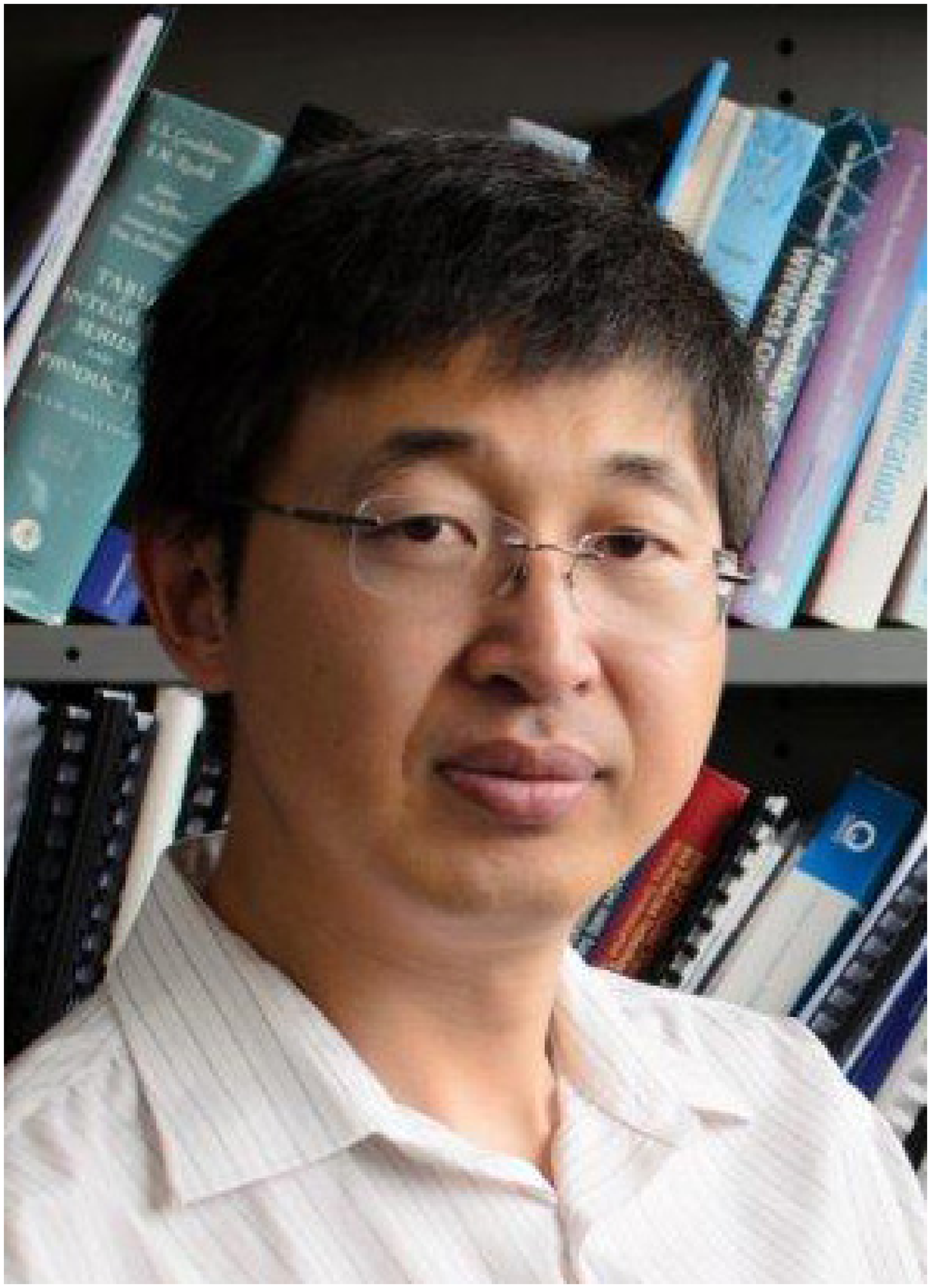}}]{Yonghui Li} (M'04-SM'09)
received his PhD degree in November 2002 from Beijing University of Aeronautics and Astronautics. From 1999 - 2003, he was affiliated with Linkair Communication Inc, where he held a position of project manager with responsibility for the design of physical layer solutions for the LAS-CDMA system. Since 2003, he has been with the Centre of Excellence in Telecommunications, the University of Sydney, Australia. He is now a Professor in School of Electrical and Information Engineering, University of Sydney. He is the recipient of the Australian Queen Elizabeth II Fellowship in 2008 and the Australian Future Fellowship in 2012.

His current research interests are in the area of wireless communications, with a particular focus on MIMO, millimeter wave communications, machine to machine communications, coding techniques and cooperative communications. He holds a number of patents granted and pending in these fields. He is now an editor for IEEE transactions on communications, IEEE transactions on vehicular technology and an executive editor for European Transactions on Telecommunications (ETT). He received the best paper awards from IEEE International Conference on Communications (ICC) 2014 and IEEE Wireless Days Conferences (WD) 2014.
\end{IEEEbiography}

\begin{IEEEbiography}[{\includegraphics[width=1in,height=1.25in,clip,keepaspectratio]{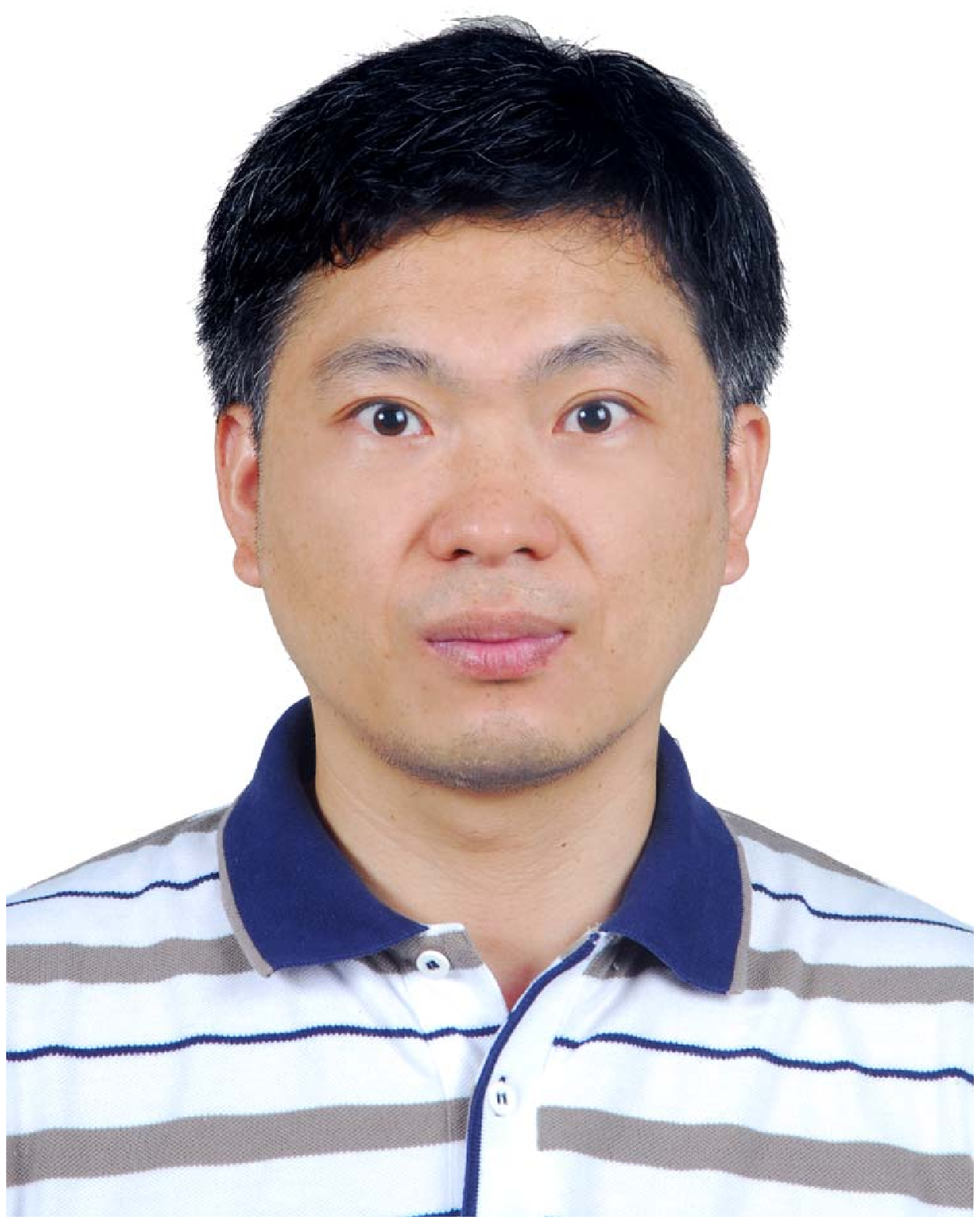}}]{Yuexing Peng} (M'06)
received the Ph.D degree in Information and communication engineering from Southeast University, Nanjing, China, in 2004. From July 2004 to December 2005, he served as senior engineer with ZTE Co.. From January 2006 to May 2008, he was a Postdoctoral Research Associate with the Telecommunication Engineering School, Beijing University of Posts and Telecommunications (BUPT). Since May 2008, he has been a faculty member with the School of Information and Communication Engineering, BUPT, where he is currently an Associate Professor. His research interests are in intelligent signal processing and machine learning and their applications to Internet of Things, wireless communications, and distributed sensing.
\end{IEEEbiography}

\begin{IEEEbiography}[{\includegraphics[width=1in,height=1.25in,clip,keepaspectratio]{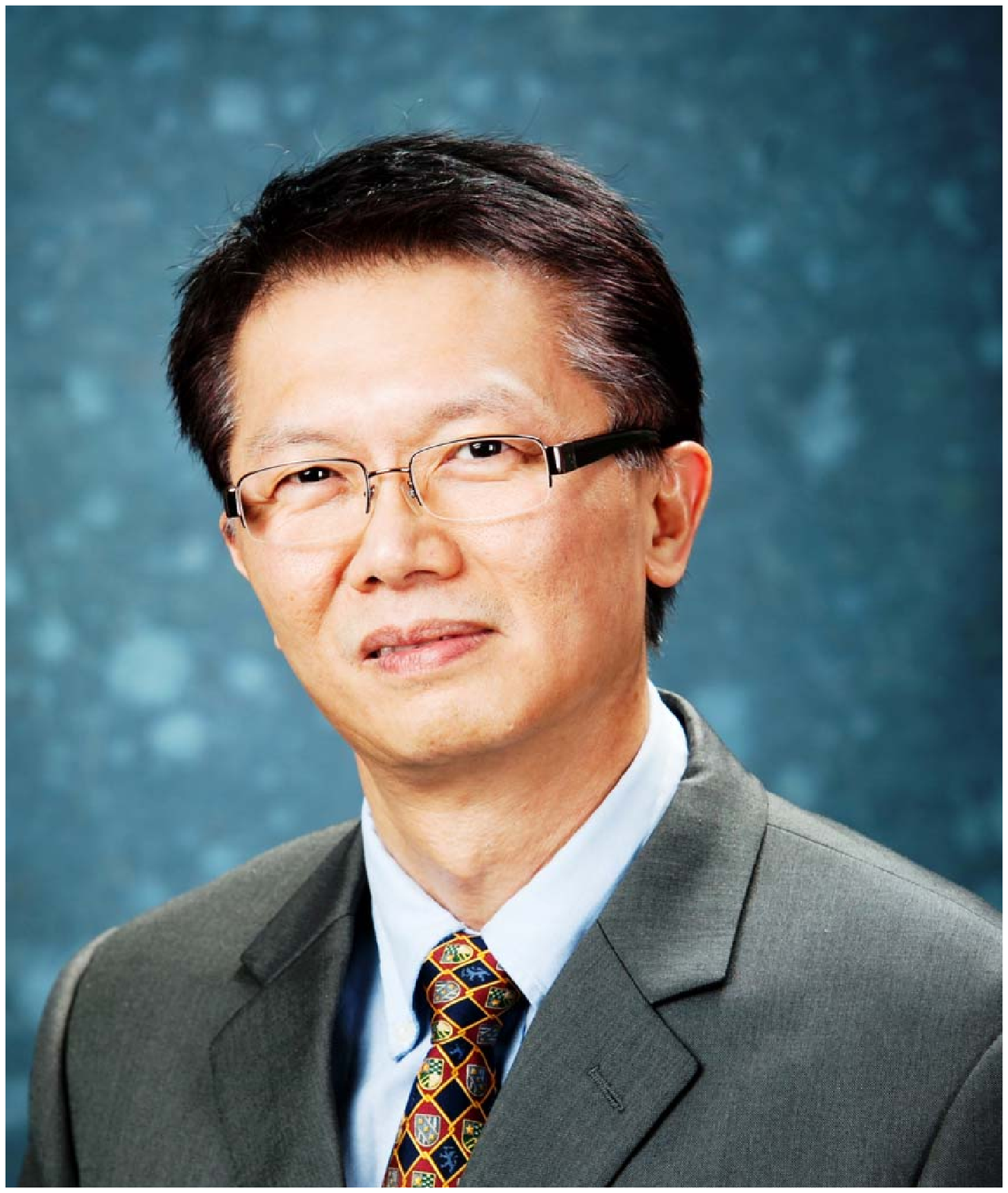}}]{Soung Chang Liew} ((M'??-SM'??-F'??))
received his  S.B., S.M., E.E., and Ph.D. degrees from the Massachusetts Institute of Technology. From 1984 to 1988, he was at the MIT Laboratory for Information and Decision Systems, where he investigated Fiber-Optic Communications Networks. From March 1988 to July 1993, he was at Bellcore (now Telcordia), New Jersey, where he engaged in Broadband Network Research. He has been a Professor at the Department of Information Engineering, the Chinese University of Hong Kong (CUHK), since 1993.  Prof. Liew is currently the Division Head of the Department of Information Engineering and a Co-Director of the Institute of Network Coding at CUHK. He is also an Adjunct Professor of Peking University and Southeast University, China.

Prof. Liew's research interests include wireless networks, Internet protocols, multimedia communications, and packet switch design. Prof. Liew's research group won the best paper awards in IEEE MASS 2004 and IEEE WLN 2004. Separately, TCP Veno, a version of TCP to improve its performance over wireless networks proposed by Prof. Liew's research group, has been incorporated into a recent release of Linux OS. In addition, Prof. Liew initiated and built the first inter-university ATM network testbed in Hong Kong in 1993. More recently, Prof. Liew's research group pioneers the concept of Physical-layer Network Coding (PNC).

Besides academic activities, Prof. Liew is active in the industry. He co-founded two technology start-ups in Internet Software and has been serving as a consultant to many companies and industrial organizations.

Prof. Liew is the holder of 11 U.S. patents and a Fellow of IEEE, IET and HKIE. He currently serves as Editor for IEEE Transactions on Wireless Communications and Ad Hoc and Sensor Wireless Networks. He is the recipient of the first Vice-Chancellor Exemplary Teaching Award in 2000 and the Research Excellence Award in 2013 at the Chinese University of Hong Kong. Publications of Prof. Liew can be found in www.ie.cuhk.edu.hk/soung.
\end{IEEEbiography}

\begin{IEEEbiography}[{\includegraphics[width=1in,height=1.25in,clip,keepaspectratio]{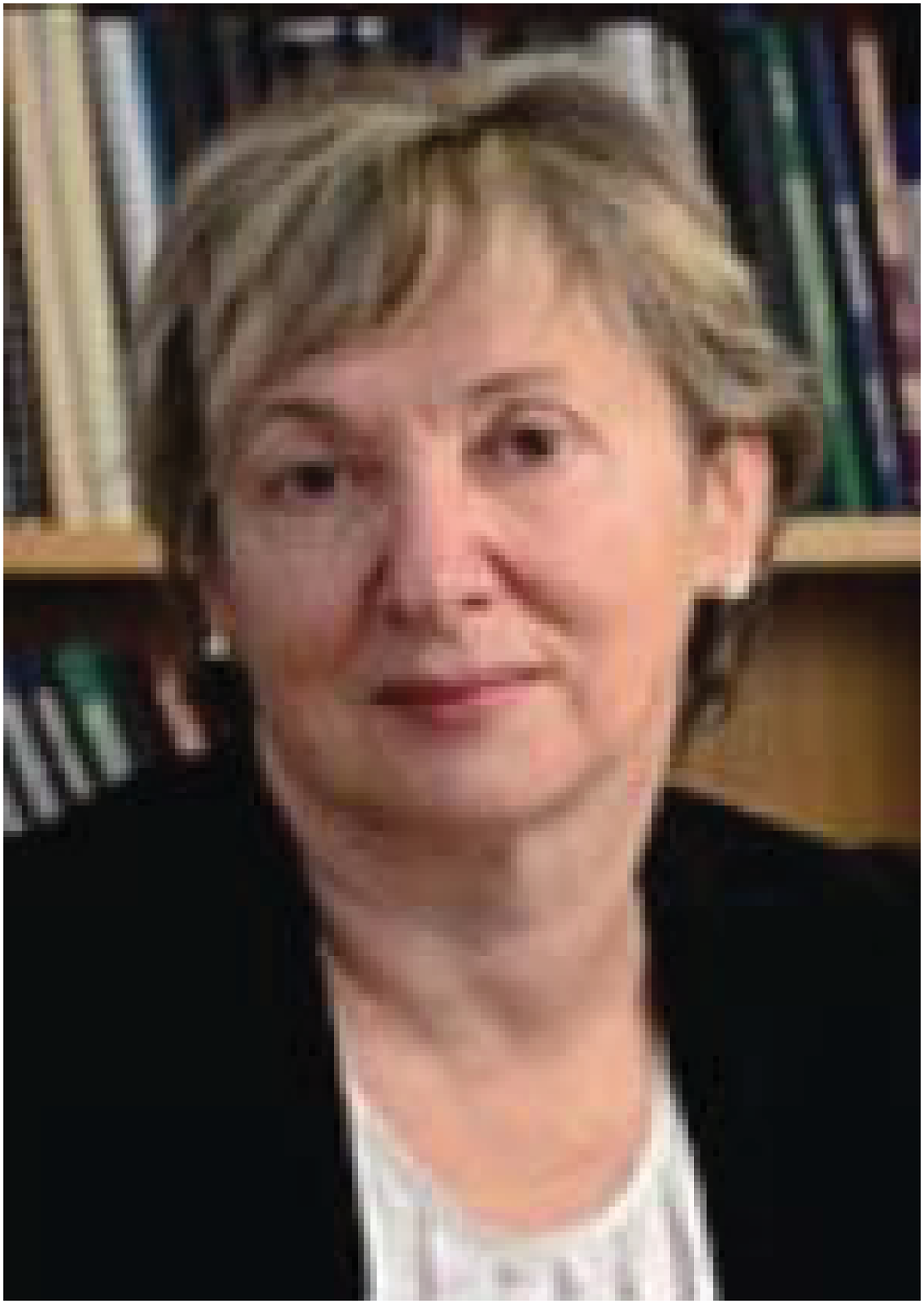}}]{Branka Vucetic} (F'03)
received the B.S.E.E., M.S.E.E., and Ph.D. degrees in electrical engineering, from the University of Belgrade, Belgrade, Yugoslavia, in 1972, 1978, and 1982, respectively. She is Laureate Professor at the University of Sydney, and Director of the Centre of Excellence in Telecommunications in The University of Sydney's School of Electrical and Information Engineering.

During her career, Prof Vucetic has held research and academic positions in Yugoslavia, Australia, UK and China. She co-authored four books and more than four hundred papers in telecommunications journals and conference proceedings. Her research interests include coding, communication theory and signal processing and their applications in wireless networks and industrial internet of things.

She is a Fellow of the Australian Academy of Technological Sciences and Engineering and a Fellow of IEEE.
\end{IEEEbiography}

\vfill
%
%




\end{document}